\documentclass[final,1p,times]{elsarticle}
\journal{arXiv}

\usepackage{graphics,amssymb,amsfonts, multirow,booktabs,amsmath,hyperref,epsfig,color,eurosym, amstext,booktabs, color,dcolumn,adjustbox,lscape}

\usepackage{graphicx}
\usepackage{lscape}
\graphicspath{{figure/}}

\begin{document}

\begin{frontmatter}

\title{Relatedness, Knowledge Diffusion, and the Evolution of Bilateral Trade}

\author[a]{Bogang Jun}
\author[b]{Aamena Alshamsi}
\author[a,c]{Jian Gao}
\author[a,cor1]{C{\'e}sar A. Hidalgo} 

\address[a]{Collective Learning Group, MIT Media Lab, Massachusetts Institute of Technology, Cambridge, MA 02139, USA}
\address[b]{Masdar Institute of Science and Technology, Abu Dhabi, UAE}
\address[c]{CompleX Lab, Web Sciences Center and Big Data Research Center, University of Electronic Science and Technology of China, Chengdu 611731, China.}
\cortext[cor1]{\emph{Email address}: hidalgo@mit.edu}

\begin{abstract}
During the last decades two important contributions have reshaped our understanding of international trade. First, countries trade more with those with whom they share history, language, and culture, suggesting that trade is limited by information frictions. Second, countries are more likely to start exporting products that are similar to their current exports, suggesting that knowledge diffusion among related industries is a key constrain shaping the diversification of exports. But does knowledge about how to export to a destination also diffuses among related products and geographic neighbors? Do countries need to learn how to trade each product to each destination? Here, we use bilateral trade data from 2000 to 2015 to show that countries are more likely to increase their exports of a product to a destination when: (i) they export related products to it, (ii) they export the same product to the neighbor of a destination, (iii) they have neighbors who export the same product to that destination. Then, we explore the magnitude of these effects for new, nascent, and experienced exporters, (exporters with and without comparative advantage in a product) and also for groups of products with different level of technological sophistication. We find that the effects of product and geographic relatedness are stronger for new exporters, and also, that the effect of product relatedness is stronger for more technologically sophisticated products. These findings support the idea that international trade is shaped by information frictions that are reduced in the presence of related products and experienced geographic neighbors.
\end{abstract}


\end{frontmatter}

\section{Introduction}
\label{intro}
{F}or more than a century, the literature on international trade explained global commerce as a consequence of differences in factor endowments \cite{Heckscher1991}, product quality, and product differentiation ~\citep{Krugman1979,Krugman1991,Anderson1979,Helpman1987}. More recent streams of literature, however, have shown that there is more to international trade than endowments, costs, and distance, since countries need to learn how to produce and export each product~\citep{Hidalgo2007,Hidalgo2015}, and also, need to overcome important information frictions to enter each export destination~\citep{Rauch1999, Rauch2001, Rauch2002, petropoulou2008,portes2005, Casella2002, Anderson2002, Garmendia2012}.

During the last two decades scholars have documented that volumes of bilateral trade decrease with the presence of borders~\citep{McCallum1995}, and increase with migrants, shared language, and social networks~\citep{Rauch2002, Rauch2001, Combes2005, Chaney2014, Bailey2017}. In fact, using the random re-allocation of the Vietnamese boat people --a population of 1.4 million Vietnamese refugees reallocated in the U.S.--, \cite{Parsons2017} showed that states who received a 10\% increase in their Vietnamese population experienced a growth in exports to Vietnam of between 4.5\% and 14\%. 

But the evidence in favor of knowledge diffusion is not only expressed on aggregated trade flows, since scholars have also shown the effects of language, social networks, and informal institutions to be larger for differentiated products \citep{Rauch1999, Rauch2001, Rauch2002}. This suggests that factors limiting knowledge and information diffusion (from social networks to language) play a more important role in the diffusion of the knowledge and information needed to exchange more sophisticated goods. 

A second stream of literature has focused on the supply side, in particular, on the process by which countries learn how to produce the products they export. This literature has shown that the ability of countries and regions to enter new export markets is limited by knowledge diffusion, since countries and regions are more likely to start exporting products when these are related to their current exports \citep{Hidalgo2007,Hidalgo2009,Hausmann2014,Boschma2013}, or when their geographic neighbors are already exporting them \citep{Bahar2014}. The importance of knowledge diffusion in the diversification of economic activities, however, is not limited only to the export of products. It has also been observed in the development of regional industries \citep{Neffke2011, Gao2017}, research activities \citep{Guevara2016}, and technologies \citep{Kogler2013, Boschma2014}, suggesting that relatedness between economic activities facilitates knowledge diffusion in general, not only in the context of international trade. 

Together, these findings have given rise to a more nuanced picture of international trade. A picture where factor endowments and transportation costs do not determine trade fully, because information frictions and knowledge diffusion determine the knowledge a country has, and hence, the products it can produce and the partners it can trade with.

Here, we contribute to this literature by combining the stream of literature on knowledge and information frictions and that on relatedness by exploring the path dependencies affecting the evolution of the network of export destinations for each product. We use more than 15 years of bilateral trade data, disaggregated into more than 1,200 products, to construct a gravity model that validates previous findings and expands them. Looking at hundreds of thousands of bilateral trade links reveals that: (i) countries are more likely to increase their exports of a product to a destination when they already export related products to it; (ii) countries are more likely to increase their exports of a product to a destination when they already export to that destination's neighbors, and (iii) countries are more likely to increase the exports of a product to a destination when their neighbors export that product to that destination. Moreover, we find that sharing a colonial past, a language, a border, or bilingual speakers (when the two countries share a language), is also associated with an increase in the volume of trade.

Yet, only some of these findings are novel. The effects of common language, shared border, and shared colonial past that we reproduce here have been documented in the past~\citep{McCallum1995, Rauch1999}. Also, we know that countries are more likely to start exporting to a destination when they export to that destination's neighbors~\citep{Chaney2014}. What is novel, are (i) the effect of relatedness on bilateral trade volumes, (ii) the effect of having a neighbor export the same product to the same destination, and (iii) the effect of bilingual speakers. In particular, we find that the effects of relatedness--the finding that countries increase their exports of a product to a destination when they already export related products to it--are especially strong. In fact, a one standard deviation increase in relatedness is associated with a 20\% increase in exports in a two year period. This effect is about 50\% larger than the effect of exporting that product to a neighbor of the target destination~\citep{Chaney2014}, and more than 170\% larger than the effect of having a neighbor export the same product to the same destination. The effect of bilingual speakers, while significant after considering all controls, are much smaller than that of sharing a language.

We also study these effects by separating exporters into new exporters, nascent, and experienced exporters, by considering as new exporters those without comparative advantage in a product. We find the effects of product relatedness, and especially those of geographic relatedness among exporters, to be stronger for new exporters than for experienced exporters. Also, we test the hypothesis that knowledge diffusion should affect more strongly products that are knowledge intensive~\citep{Rauch1999, Rauch2001, Rauch2002} by dividing products into the five technological categories suggested by Lall~\citep{Lall2000}: primary, resource-based manufactures, low-tech, medium-tech, and high-tech products. We find that exporting related products has a stronger effect on the increase of exports for technological sophisticated products, suggesting that knowledge diffusion is more important for knowledge intense products. Surprisingly, we find no effect of technological sophistication on both of our measures of regional relatedness. That is, neither exporting to a neighbor, nor having a neighbor export the same product, appears to have an effect on future exports that either increases or decreases with technological sophistication. Also, we find that sharing a language and a colonial past has a larger effect on the increase of exports for more technologically sophisticated products, providing further evidence that the effects are due to information and knowledge frictions. Moreover, we find the negative effect of distance--which are correlated with social network connections~\citep{Breschi2009,Singh2005}--to be larger for technologically sophisticated products. These findings support the idea that establishing trade relationships requires overcoming information and knowledge frictions, and that product and geographic relatedness help reduce these frictions.

\section{Data}
\label{data}

We use bilateral trade data from 2000 to 2015 from MIT's Observatory of Economic Complexity~\citep{Simoes2011}. The data is disaggregated into the Harmonized System (HS rev 1992, four-digit level) and consists of imports and exports between countries. Because both exporter and importer report their trade information, we clean the data by comparing the data reported by exporters and importers following the work of \cite{Feenstra2005}. Also, we exclude countries that have population less than 1.2 million or have a trade volume in 2008 that is below 1 billion in US dollar. Also, we exclude data from Iraq, Chad and Macau.

Macroeconomic data (GDP at market prices in current US dollar and population) comes from the World Bank's World Development Indicators. Data on geographical and cultural distance (shared language, physical distance between most populated cities, sharing a border, and shared colonial past) comes from GEODIST data from CEPII \citep{Mayer2011}. For language proximity, we use one of the global language networks of \cite{Ronen2014}: the one considering the number of books translated from one language to another as a proxy for the number of translators, or bilingual speakers, between two languages. 

\section{Results}
Does relatedness among products or geographic neighbors help facilitate the knowledge flows needed to increase bilateral trade flows?

\begin{figure}[!t]
\centering
\includegraphics[width=1\linewidth]{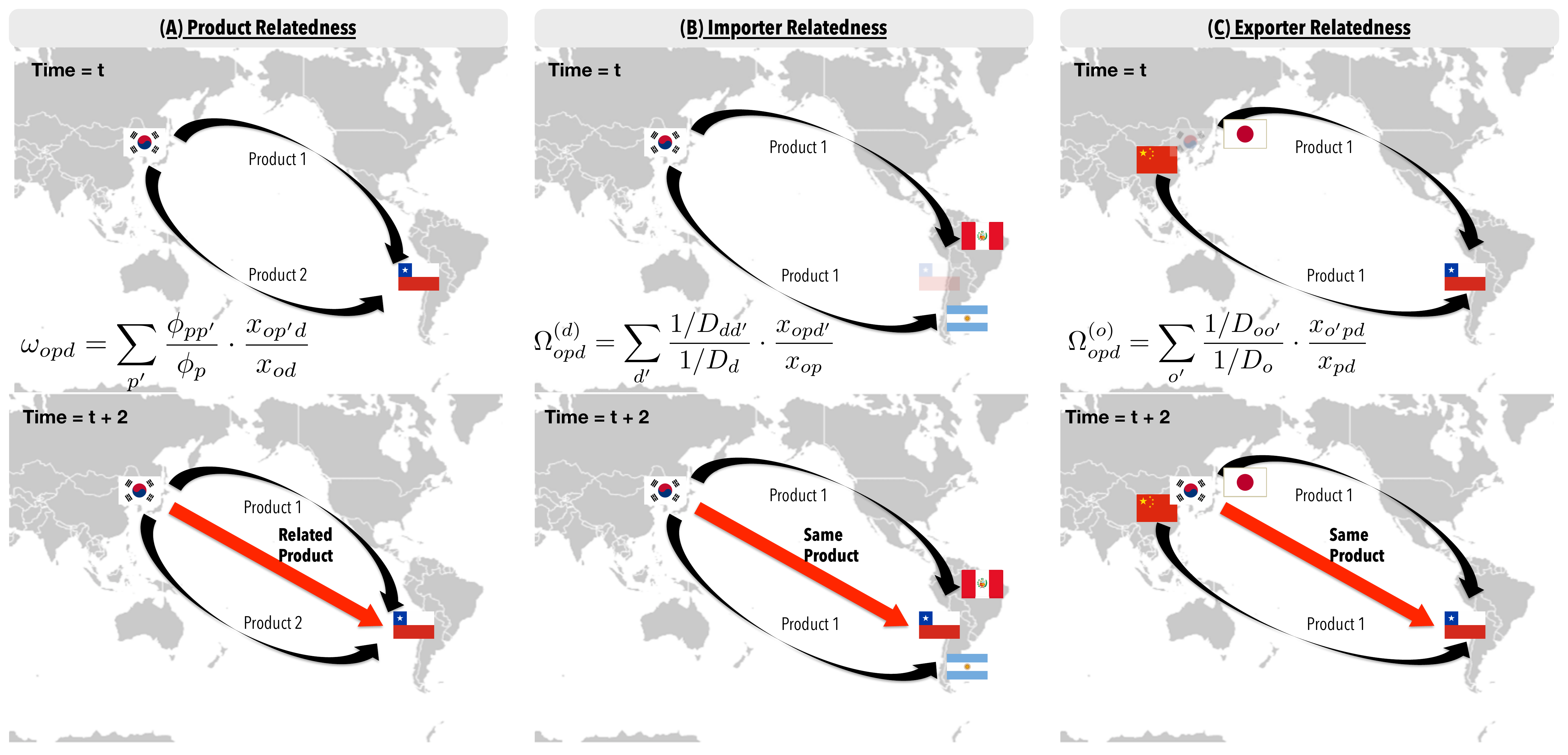}
\caption{Relatedness among products, exporters, and importers. (A) Product Relatedness: the similarity between a product and the other products that a country already exports to a destination, (B) Importer Relatedness: the fraction of the geographic neighbors of a country that import a product from the same origin, and (C) Exporter Relatedness: the fraction of neighbors of a country that export a product to the same destination.}
\label{learningChannels}
\end{figure}

To explore this question we introduce three measures of relatedness. We use these to estimate: (i) the similarity between a product and the other products that a country already exports to a destination (Product Relatedness), (ii) the fraction of the geographic neighbors of a country that import a product from the same origin (Importer Relatedness), and (iii) the fraction of neighbors of a country that export a product to the same destination (Exporter Relatedness). Product Relatedness should help us capture information about knowledge flows between products (which range from knowledge flows among industries to knowledge flows among product lines within a firm). Figure~\ref{learningChannels}(A) illustrates Product Relatedness in the context of Korea and Chile. In the example, Korea exports Products I and II to Chile (Shirts and Pants), and this may affect the future exports of Product III (Coats) to Chile, when Product III (Coats) is highly related to Products I and II (Shirts and Pants). Our hypothesis is that knowledge flows should be larger among related products, and hence, exports should increase faster when a country exports related products to a destination. 

Importer Relatedness helps us capture knowledge flows on how to: (i) import a product from the same origin than a neighbor, or (ii) export to a neighbor of a current destination. In the example of Figure~\ref{learningChannels}(B), Korea exports Product I (Shirts) to Peru and Argentina and that may affect the future volume of exports of Product I (Shirts) to Chile (who is a geographic neighbor of Peru and Argentina). Here, knowledge on how to import from an origin should be flowing among neighboring importers, or knowledge on how to export to the neighbor's of a country's destinations should be flowing within the exporter. 

Exporter Relatedness captures (i) knowledge flows among neighboring exporters on how to export to a destination, or (ii) knowledge flows on how to import from a neighbor of a country from where you currently import. In the example of Figure~\ref{learningChannels}(C), Chile imports Product I (Shirts) from China and Japan, and that may affect the future volume of exports of Product I (Shirts) from Korea (which is a neighbor of the places from where Chile is currently importing Product I). This would be a knowledge flow on how to export to a destination among neighboring exporters, or a knowledge flow within an importer, of how to import from a neighbor of a current origin.  

Mathematically, we can construct the three measures of relatedness using a similar formula. The formula  is a weighted average of the number of neighbors, or related products, that already have an active trade relationships. In the case of similarity between products, weighs are the proximity between products $p$ and $p'$, $\phi_{pp'}$. $\phi_{pp'}$ is the minimum of the conditional probability that two products are co-exported by multiple countries (see ~\cite{Hidalgo2007} and Appendix A). $\phi_{pp'}=1$ means products $p$ and $p'$ are always co-exported and $\phi_{pp'}=0$ means no country exports both products. In the case of geographic neighbors, of both an exporter or an importer, weights are given by the inverse of geographic distance ($1/D_{dd'}$ and $1/D_{oo'}$), where $D_{dd'}$ is the distance in kilometers between the most populated cities in countries $d$ and $d'$. 

Formally, let $x_{opd}$ be a matrix summarizing the trade flow in US dollars of product $p$ from exporter $o$ to destination $d$. Then, Product Relatedness is given by:  

\begin{equation}
        \omega_{opd}= \sum_{p'}\frac{\phi_{pp'}}{\phi_p} \cdot \frac{x_{op'd}}{x_{od}}
    \label{commercialIndustryDensity}
\end{equation}
where $x_{od}$ is the volume of trade between countries $o$ and $d$ ($x_{od}=\sum_{p}x_{opd}$) and $\phi_{p}=\sum_{p'}\phi_{pp'}$.\\ 

Similarly, Importer Relatedness is given by:
\begin{equation}
    \Omega^{(d)}_{opd} = \sum_{d'} \frac{1/D_{dd'}}{1/D_d} \cdot \frac{x_{opd'}}{x_{op}} ,
    \label{commercialRegionalDensity1}
\end{equation}
where  $x_{op}$ is the volume of trade of product $p$ from origin country $o$ ($x_{op}=\sum_{d}x_{opd}$), $D_{dd'}$ is the geographic distance between destination country $d$ and its neighbors $d'$, and $1/D_d = \sum_{d'}1/D_{dd'}$.\\ 

Finally, Exporter Relatedness is given by: 
\begin{equation}
    \Omega^{(o)}_{opd} = \sum_{o'} \frac{1/D_{oo'}}{1/D_o} \cdot \frac{x_{o'pd}}{x_{pd}}  ,
    \label{commercialRegionalDensity2}
\end{equation}
where $x_{pd}$ is the volume of trade of product $p$ to destination country $d$ ($x_{pd}=\sum_{o}x_{opd}$), $D_{oo'}$ is the geographic distance between origin country $o$ and its neighbors $o'$, and $1/D_o = \sum_{o'}1/D_{oo'}$.\\ 

Next, we use these three measures of relatedness, together with data on common cultural and geographic factors, to construct an extended gravity model to study the marginal contribution of product, importer, and exporter relatedness, and of shared languages, borders, and colonial past, to the growth of future exports. Our model predicts bilateral trade in a product in two years time while controlling for: (i) initial trade in that product between the same trade partners, (ii) total exports of the product by the exporter, (iii) total imports of the product by the importer, (iv-vii) the GDP per capita and population of exporters and importers, and (viii) their geographic distance. Formally, our model is given by Equation~\ref{regressionEquation}:

\begin{equation}
 \begin{split}
    x^{t+2}_{opd} &= \beta_{0} + \beta_{1}\omega^{t}_{opd} + \beta_{2}\Omega^{(d)t}_{opd} + \beta_{3}\Omega^{(o)t}_{opd}\\
    + \beta&_{4} x^t_{opd} + \beta_{5} x^t_{op} + \beta_{6} x^t_{pd} + \beta_{7} D_{od}  \\ 
    + \beta&_{8} gdp ^{t}_{o} + \beta_{9} gdp ^{t}_{d}  + \beta_{10}Population^{t}_{o}+ \beta_{11}Population^{t}_{d}\\
    + \beta&_{12} Border_{od}  + \beta_{13} Colony_{od}+ \beta_{14}Language_{od}\\
    +\beta&_{15}Lang.Proximity_{od}\\
    + \varepsilon&^t_{opd}
     \end{split}
\label{regressionEquation}
\end{equation} where the dependent variable, $x^{t+2}_{opd}$, represents the volume of trade in (US dollar) of product $p$ from exporter $o$ to destination $d$ in year $t+2$. Our main variables of interest are our three measures of relatedness: Product Relatedness $\omega^{t}_{opd}$, Importer Relatedness $\Omega^{(d)t}_{opd}$, Exporter Relatedness $\Omega^{(o)t}_{opd}$, and shared border ($Border_{od}$), shared language ($Language_{od}$), language proximity (number of bilingual speakers $Lang.Proximity_{od}$), and shared colonial past ($Colony_{od}$). $Border_{od}$, $Language_{od}$, $Colony_{od}$ are binary (dummy) variables (0 or 1). The other factors in the model $gdp$ per capita, population, and distance ($D_{od}$), are standard gravity controls~\citep{Tinbergen1962,Poyhonen1963}. Finally, by incorporating the total volume of exports of a country ($x_{op}$), the total imports of a destination ($x_{pd}$), and the present day trade flow for each product between an origin and a destination ($x_{opd}$), we capture the effects of our variable of interest in the change in trade experience in the subsequent two years. In Equation~\ref{regressionEquation}, we make all variables comparable (except binary variables) by standardizing them by subtracting their means and dividing them by their standard deviations. (Please see Appendix B and C for summary statistics and correlation among variables)

\begin{table}[t] 
\centering 
  \caption{Bilateral trade volume after two years for periods 2000-2006 (pre-financial crisis), 2007-2012 (crisis period) and 2012-2015 (recovery period)} 
  \label{main1} 
    \begin{adjustbox}{width=0.59\textwidth,center}
\begin{tabular}{@{\extracolsep{5pt}}lD{.}{.}{-3} D{.}{.}{-3} D{.}{.}{-3}} 
\\[-1.8ex]\hline 
\hline \\[-1.8ex] 
 & \multicolumn{3}{c}{\textit{Dependent variable}:~log $x^{t+2}_{opd}$} \\ 
\cline{2-4} 
\\[-1.8ex] & \multicolumn{1}{c}{\textit{2000-2006}} & \multicolumn{1}{c}{\textit{2007-2012}}& \multicolumn{1}{c}{\textit{2012-2015}}\\ 
\hline \\[-1.8ex] 
  $\omega^t_{opd}$ & 0.209^{***} & 0.180^{***}  & 0.152^{***}\\ 
  & (0.001)  & (0.001) & (0.001)\\ 
  $\Omega^{(d)}_{opd}$ & 0.143^{***} & 0.143^{***}  & 0.151^{***} \\ 
  & (0.001) & (0.001) & (0.001)\\ 
  $\Omega^{(o)}_{opd}$ & 0.077^{***} & 0.105^{***}  & 0.107^{***} \\ 
  & (0.001) & (0.001) & (0.001)\\ 
  log $x^t_{opd}$ & 1.371^{***} & 1.600^{***}  & 1.769^{***} \\ 
  & (0.001) & (0.001) & (0.001)\\ 
  log $x^t_{op}$ & 0.961^{***}& 0.972^{***}  & 0.915^{***} \\ 
  & (0.001) & (0.002)  & (0.002) \\ 
  log $x^t_{pd}$  & 0.529^{***} & 0.421^{***}  & 0.406^{***} \\ 
   & (0.001)  & (0.001)  & (0.002)\\ 
  log $Distance$  & -0.485^{***} & -0.419^{***}  & -0.432^{***}\\ 
  & (0.001) & (0.001)  & (0.004)\\ 
  log $gdp^t_o$ &0.165^{***} & 0.198^{***} &0.203^{***} \\ 
  & (0.001) & (0.001) & (0.001) \\ 
  log $gdp^t_d$ &  0.226^{***}&  0.216^{***} &  0.281^{***}\\ 
  &  (0.001) &  (0.001) &  (0.001)\\ 
  log $Population_o$ &  0.472^{***}&  0.456^{***} &  0.455^{***} \\ 
  &  (0.001)&  (0.001) &  (0.002)\\ 
  log $Population_d$  &  0.344^{***}&  0.367^{***} &  0.338^{***}\\ 
   & (0.001) &  (0.001) &  (0.001) \\ 
  $Border_{od}$ &  0.712^{***}&  0.717^{***} & 0.611^{***}\\ 
  &  (0.003)&  (0.004) &  (0.005)\\ 
  $Colony_{od}$ &  0.052^{***} &  0.127^{***} &  0.193^{***}\\ 
   &  (0.003)&  (0.004) & (0.005) \\ 
  $Language_{od}$ &  0.545^{***} &  0.441^{***} &  0.334^{***}\\ 
  &  (0.002) &  (0.003) &  (0.003) \\ 
  log$Lang. Proximity_{od}$  &  0.032^{***}&  0.027^{***} &  -0.001 \\ 
  &  (0.001)&  (0.001) &  (0.001) \\ 
  Constant  & 9.653^{***}     & 9.830^{***}  & 9.793^{***}\\ 
  & (0.001)   & (0.001) & (0.001)  \\ 
 \hline \\[-1.8ex] 
Observations  & \multicolumn{1}{c}{10,911,584} & \multicolumn{1}{c}{7,591,489}  & \multicolumn{1}{c}{5,332,257}\\ 
Adjusted R$^{2}$  & \multicolumn{1}{c}{0.494} & \multicolumn{1}{c}{0.516} & \multicolumn{1}{c}{0.558}\\ 
Residual Std. Error & \multicolumn{1}{c}{2.568} & \multicolumn{1}{c}{2.637} & \multicolumn{1}{c}{2.529} \\ 
\hline 
\hline \\[-1.8ex] 
\textit{Note:}  & \multicolumn{2}{r}{$^{*}$p$<$0.1; $^{**}$p$<$0.05; $^{***}$p$<$0.01} \\ 
\end{tabular} 
\end{adjustbox}
\end{table} 

Table ~\ref{main1} shows our main results divided into three periods: 2000-2006 (pre-financial crisis), 2007-2012 (crisis period), and 2012-2015 (recovery period). Since our result are qualitatively the same for all of these periods we will describe them in unison. (See detailed results, correlation table and summary statistics in Appendix B and C)

First, we find that the three relatedness variables correlate positively with future bilateral trade. This means that a country is likely to experience an increase in their exports of product $p$ to a destination $d$ when: (i) the country is exporting related products to that destination, (ii) it is exporting the same product to the neighbors of a destination (confirming \cite{Chaney2014}), and (iii) it has neighbors that are already exporting the same product to that destination. This extends \cite{Bahar2014}, who showed that having geographic neighbors increases the probability of exporting a new product, since \cite{Bahar2014} did not look at individual export destinations (they aggregate across all destinations). Our findings, therefore, complement \cite{Bahar2014} by showing that having neighbors that export a product does not only increase the total volume of exports, but the volume of exports to the same destinations that the neighbors were exporting to. 

When comparing the effects of product and geographical relatedness (variables are standardized), we find that the role of Product Relatedness ($\omega^t_{opd}$) is on average the largest, while that of Exporter Relatedness ($\Omega^{o}_{opd}$) is the smallest. In addition to these, we find strong and positive effects for the role of shared borders, colonial past, shared language, and to a lesser extent, number of translations (language proximity). Other standard gravity factors (distance, GDP per capita, and population) behave as expected.

Together, the finding that relatedness among products, the presence of knowledge among geographic neighbors, language, colonial history, shared borders, and language proximity, all have a positive and significant effect in the increase of trade flows for particular products and countries, are evidence in support of the notion that knowledge on how to trade a specific product between a specific pair of countries needs to flow for that trade to be materialized. If this hypothesis is correct, we should also be able to study the varying importance of knowledge flows for new and experienced exporters (exporters with or without comparative advantage), and also, for products with different levels of technological sophistication.

\begin{table}[t] \centering 
  \caption{Bilateral trade volume after two years for new, nascent, and experienced exporters} 
  \label{main2} 
    \begin{adjustbox}{width=0.59\textwidth,center}
\begin{tabular}{@{\extracolsep{5pt}}lD{.}{.}{-3} D{.}{.}{-3} D{.}{.}{-3} } 
\\[-1.8ex]\hline 
\hline \\[-1.8ex] 
 & \multicolumn{3}{c}{\textit{Dependent variable}:~log $x^{t+2}_{opd}$} \\ 
\cline{2-4} 
\\[-1.8ex] & \multicolumn{1}{c}{New Exporter} & \multicolumn{1}{c}{Nascent Exporter} & \multicolumn{1}{c}{Experienced Exporter}\\ 
\hline \\[-1.8ex] 
 $\omega^t_{opd}$ & 0.194^{***} & 0.140^{***} & 0.184^{***} \\ 
  & (0.003) & (0.004) & (0.001) \\ 
  $\Omega^{(d)}_{opd}$ & 0.149^{***} & 0.122^{***} & 0.122^{***} \\ 
  & (0.003) & (0.005) & (0.001) \\ 
  $\Omega^{(o)}_{opd}$ & 0.111^{***} & 0.078^{***} & 0.069^{***} \\ 
  & (0.003) & (0.005) & (0.001) \\ 
  log $x^t_{opd}$ & 0.474^{***} & 1.410^{***} & 1.484^{***} \\ 
  & (0.003) & (0.005) & (0.001) \\ 
  log $x^t_{op}$ & 0.964^{***} & 0.684^{***} & 0.804^{***} \\ 
  & (0.005) & (0.006) & (0.001) \\ 
  log $x^t_{pd}$ & 0.525^{***} & 0.518^{***} & 0.553^{***} \\ 
  & (0.005) & (0.008) & (0.001) \\ 
  log $Distance$ & -0.532^{***} & -0.504^{***} & -0.454^{***} \\ 
  & (0.004) & (0.006) & (0.001) \\ 
  log $gdp^t_o$ & 0.235^{***} & 0.128^{***} & 0.100^{***} \\ 
  & (0.004) & (0.005) & (0.001) \\ 
  log $gdp^t_d$ & 0.010^{**} & 0.045^{***} & 0.264^{***} \\ 
  & (0.005) & (0.007) & (0.001) \\ 
  log $Population_o$ & 0.331^{***} & 0.388^{***} & 0.446^{***} \\ 
  & (0.003) & (0.005) & (0.001) \\ 
  log $Population_d$ & 0.190^{***} & 0.213^{***} & 0.362^{***} \\ 
  & (0.004) & (0.006) & (0.001) \\ 
  $Border_{od}$ & 0.651^{***} & 0.817^{***} & 0.713^{***} \\ 
  & (0.009) & (0.016) & (0.004) \\ 
  $Colony_{od}$ & 0.218^{***} & 0.259^{***} & 0.097^{***} \\ 
  & (0.012) & (0.020) & (0.004) \\ 
  $Language_{od}$ & 0.557^{***} & 0.782^{***} & 0.481^{***} \\ 
  & (0.007) & (0.011) & (0.003) \\ 
  log$Lang. Proximity_{od}$ & 0.008^{***} & 0.016^{***} & 0.029^{***} \\ 
  & (0.003) & (0.004) & (0.001) \\ 
  Constant & 7.776^{***} & 9.213^{***} & 10.124^{***} \\ 
  & (0.004) & (0.004) & (0.001) \\ 
 \hline \\[-1.8ex] 
Observations & \multicolumn{1}{c}{922,092} & \multicolumn{1}{c}{463,388} & \multicolumn{1}{c}{8,045,262} \\ 
Adjusted R$^{2}$ & \multicolumn{1}{c}{0.281} & \multicolumn{1}{c}{0.456} & \multicolumn{1}{c}{0.526} \\ 
Residual Std. Error & \multicolumn{1}{c}{2.641} & \multicolumn{1}{c}{2.535} & \multicolumn{1}{c}{2.487} \\ 
\hline 
\hline \\[-1.8ex] 
\textit{Note:}  & \multicolumn{3}{r}{$^{*}$p$<$0.1; $^{**}$p$<$0.05; $^{***}$p$<$0.01} \\ 
\end{tabular} 
\end{adjustbox}
\end{table}

Next, we test the effects of the exporters' level of competitiveness in the diffusion of the information needed to trade by dividing exporters of each product into new, nascent, and experienced. We do this by calculating the revealed comparative advantage (RCA) of each exporter in each product. RCA is the ratio between the exports of a country in a product, and the exports that are expected based on a country's total export market and the size of the global market for that product. We classify as new exporters all countries with an RCA below 0.2 in a product (countries that export less than 20\% of what they are expected to export by chance). We classify as nascent exporters, all countries with an RCA between 0.2 and 1. We classify as the experienced exporters of a product, all countries that have revealed comparative advantage in it (RCA $>1$).

Table~\ref{main2} divides country-product pairs into new, nascent, and experienced exporters. The results are consistent with those presented in Table~\ref{main1}, but also, reveal two important distinctions. First, the effects of product and geographic relatedness, especially Exporter Relatedness, are stronger for new exporters, suggesting that knowledge and information frictions impose larger constraints for countries that are not experienced in the export of a product. Second, the overall explanatory power of the model is considerably larger for experienced exporters ($R^2\approx53\%$ vs $R^2\approx46\%$ for nascent exporters and $R^2\approx28\%$ for new exporters. These are large differences, even considering that the sample sizes are not the same). This suggests that inexperienced exporters face more uncertainty (less predictable because of lower R-square), and hence, benefit more from relatedness (higher relatedness coefficients).

Finally, we explore the interaction between our three measures of relatedness and the technological sophistication of products using \cite{Lall2000}'s five technological categories: primary, resource-based manufactures, low-tech, medium-tech, and high-tech products. Since Lall's classification is based on the 3-digit Standard International Trade Classification (SITC-3) rev 2, we match products to our data using the conversion table provided by the UN Trade Statistics site \footnote{Available at https://unstats.un.org/unsd/trade/classifications/correspondence-tables.asp}. Following \cite{Lall2000}, we also exclude ``special transactions'' such as electric current, cinema film, printed matter, fold, coins, and pets.

\begin{figure}[!t]
\centering
\includegraphics[width=0.8\linewidth]{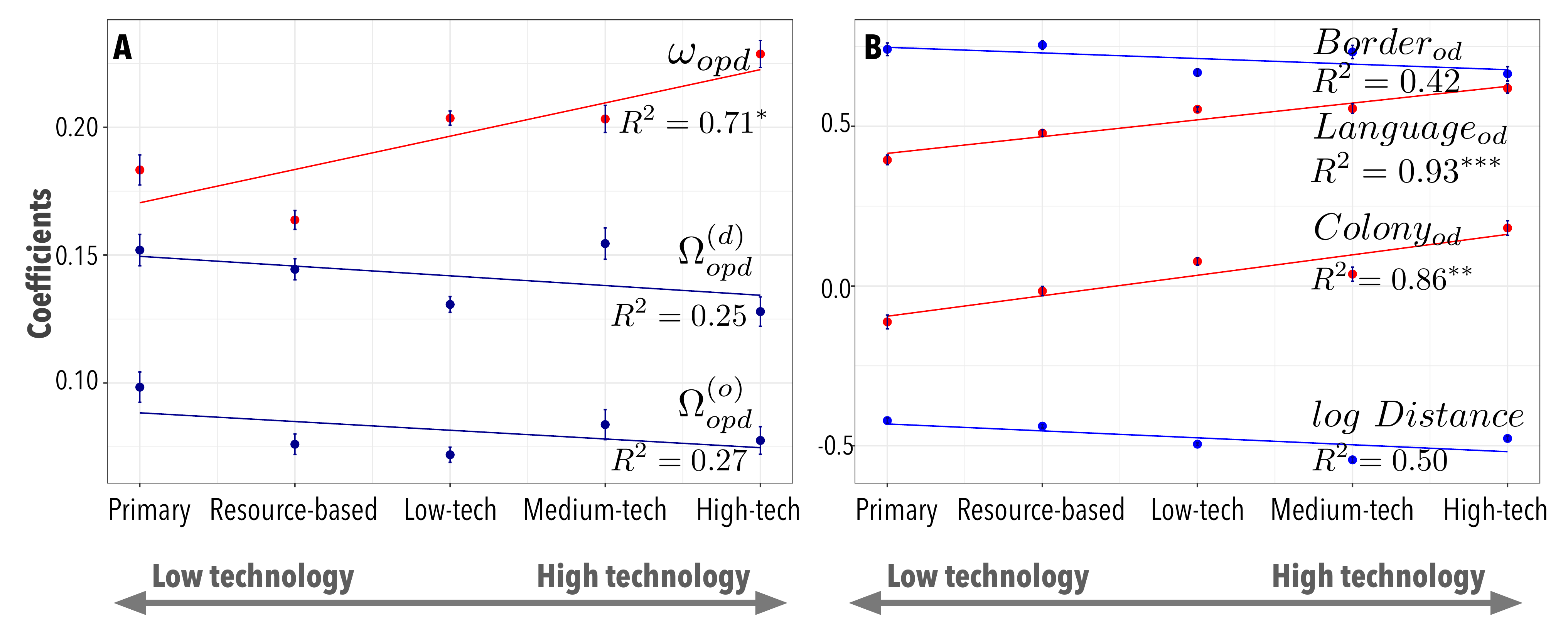}
\caption{Coefficients of variables by the technological sophistication of products; (A) Coefficients of $\omega_{opd}$, $\Omega^{(d)}_{opd}$, and $\Omega^{(o)}_{opd}$, (B) Coefficients of $Border_{od}$, $Language_{od}$, and $Colony_{od}$, and (C) Coefficients of log~$lang.Proximity$  and log~$Distance$. The fitted lines in red are statistically significant, while the lines in blue are not statistically significant. $^{*}$p$<$0.1; $^{**}$p$<$0.05; $^{***}$p$<$0.01}
\label{LallClassification}
\end{figure}

We present the coefficients for all of the variables we had in the previous model in the Appendix E (Table A15) and summarize the main results graphically, by plotting the coefficients and their errors as a function of the technological sophistication of products in Figure~\ref{LallClassification}. Trends that increase significantly with technological sophistication ($p<0.1$) are presented in red, non-significant trends are shown in blue.

Figure~\ref{LallClassification} A shows that the effect of Product Relatedness, but not that of Importer Relatedness or Exporter Relatedness, increases with technological sophistication. This suggests that product relatedness captures channels of knowledge and information flow that are relevant for the export of sophisticated products. Also, Figure~\ref{LallClassification} B shows that the effect of sharing a language and a colonial past, but not those of sharing a border, are larger for more technologically sophisticated products. Once again, this reiterates the idea that borders and geographic distance affect knowledge flows by limiting social interactions \citep{Singh2005,Breschi2009}, so we do not see much of a geographic effect once we take cultural and linguistic similarity into account. The negative effect of distance is slightly larger for technologically sophisticated products, but the effect is not strong enough to be significant. Together, these findings support the idea that trade is driven partly by the diffusion of knowledge and information on how to export each product to each destination. 

\section{Discussion}

During the last decades two ideas have re-framed our understanding of international trade. The first idea is that information and knowledge frictions, not just differences in transportation costs, factor endowments, and differences in productivity, shape global trade. The second idea is that countries need to learn how to produce the products they export, and hence, evolve their productive structures in a path dependent manner that is constrained by knowledge flows. Here, we use bilateral trade data, together with various measures of economic size, culture, and geographic proximity, to put these two ideas together. Our findings confirm many existing theories involving the role of language, and culture, but also, add to the body of knowledge by showing that relatedness among products and countries shape future trade volumes. In particular, we showed that relatedness among products, and among geographic neighbors, explains a substantial fraction of future bilateral trade: trade volumes increase when countries export related products to a destination, but also, when they share neighbors who export to that destination, or when they are already exporting to a destinations neighbors. When comparing these three forms of relatedness, we found that relatedness among products is the strongest, suggesting that there may be product or industry specific learning channels that play an important role in the diffusion of the knowledge needed to establish or increase trade relationships. Moreover, we found the effects of relatedness to be stronger for new exporters, and the effects of product relatedness to be stronger for more technologically sophisticated products. These additional considerations support the idea that the presence of related activities facilitates the knowledge flows that countries need to learn how to produce and export products to specific destinations.  

Yet, our results leave unanswered many questions about the mechanisms underlying these knowledge and information flows. The two channels we observed among geographic neighbors (Importer and Exporter Relatedness) could further be disaggregated into four channels: the knowledge flows among neighboring importers or exporters, or the knowledge flows within an exporter or within an importer. From there, we may be able to start learning about the specific mechanisms that underlie each of these knowledge flows. Also, the interpretation of relatedness has a similar problem. On the one hand, one cannot know if the flow of knowledge is among product lines, within the same firm or industry, among industries, or the result of spin-offs, foreign direct investment, or migrations. 

Nevertheless, our results do provide some light in the long quest to understand how social networks, culture, and knowledge flows, shape international trade. They tell us that product relatedness plays an important role since the size of its effect is larger than the one observed among geographic neighbors. This suggests that looking at knowledge flows among product lines, and among industries, should be an avenue of inquiry for improving our understanding of the social and economic forces that shape global trade.

\section*{Acknowledgement}
We thank Mauricio (Pacha) Vargas and Alex Simoes for help with the data. We also thank Cristian Jara Figueroa, Fl{\'a}vio Pinheiro, Tarik Roukny, and Dogyoon Song for helpful comments. This project is funded by the MIT Skoltech Program and by the Cooperative Agreement between the Masdar Institute of Science and Technology and the MIT Media Lab Consortia.


\clearpage
\linespread{1.5}
\biboptions{authoryear}

\appendix
\clearpage

\setcounter{figure}{0}
\setcounter{table}{0}
\setcounter{equation}{0}
\setcounter{page}{1}
\renewcommand{\thefigure}{A\arabic{figure}}
\renewcommand{\thetable}{A\arabic{table}}
\renewcommand\theequation{A\arabic{equation}}

\section*{Appendix A. Building a product space for 2000-2015}
\label{appendixA}
To calculate the $\omega_{opd}$, we need firstly build a product space. We define the product space by looking at all proximity measures between products \citep{Hidalgo2007} after aggregating all the data that covers from 2000 to 2015. To capture the significant trade flow, we calculate the revealed comparative advantage (RCA) following \cite{Balassa1965}: 
\begin{equation}
    RCA_{o,i} = \left.{\frac{x_{o,i}}{\sum_{i}x_{o,i}}}\bigg/ \frac{\sum_{o}x_{o,i}}{\sum_{o,i}x_{o,i}}\right.
    \label{Eq:RCA}
\end{equation}

Based on the result of RCA, we measure the proximity between product by calculating $\phi_{i,j}$ between product $i$ and $j$ \citep{Hidalgo2007}.
\begin{equation}
    \phi_{i,j} = \left.min \Big\{P(RCA_{i}|RCA_{j}, P(RCA_{j}|RCA_{i})) \Big\}\right.
    \label{Eq:industrySimilarity}
\end{equation}

\begin{figure}[!b]
  \centering
  \includegraphics[width=1\textwidth]{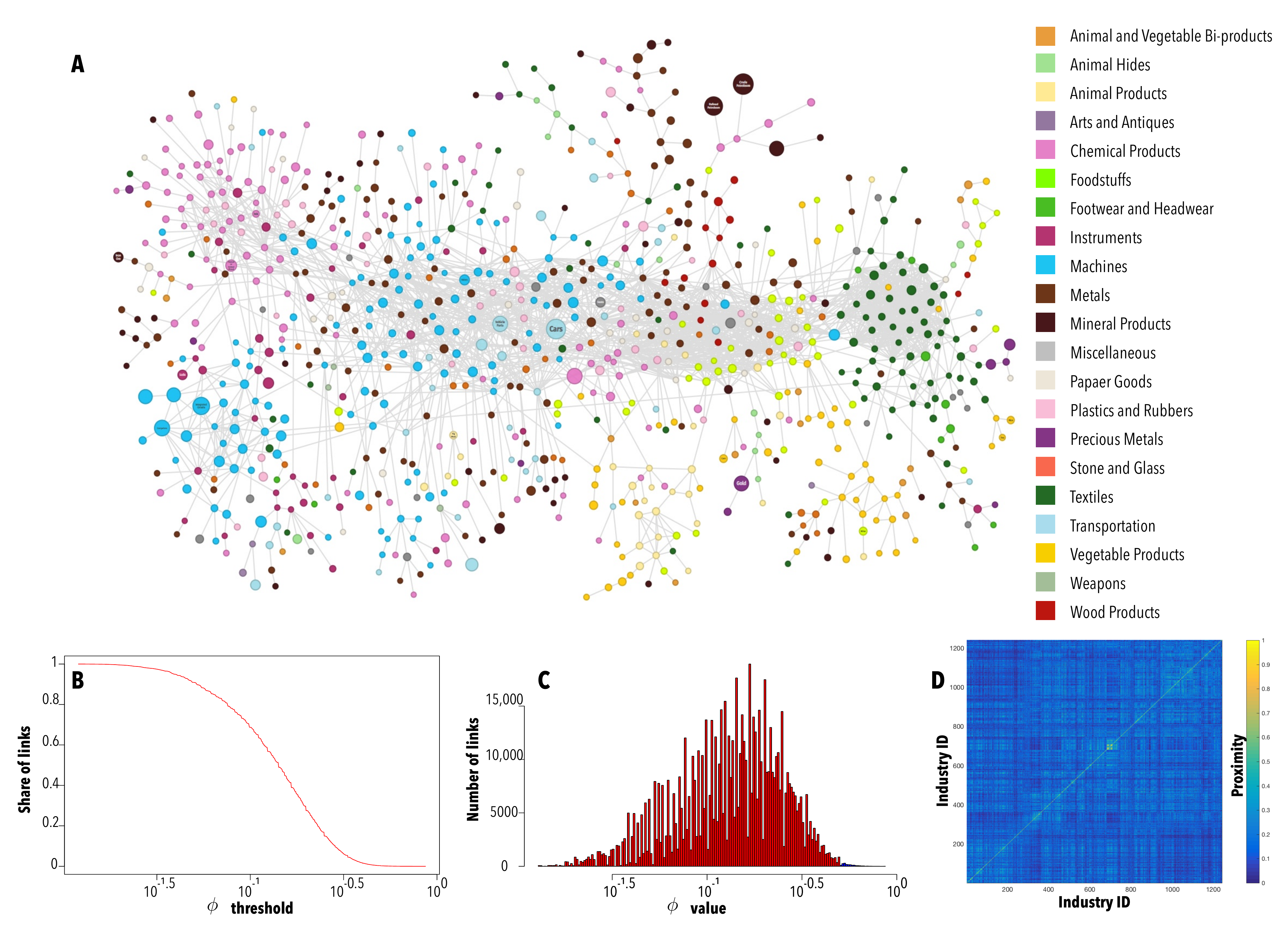}
  \caption{Product space over 2000 to 2015: (A) Network representation of product space, (B) Cumulative distribution of proximity values, (C) Density distribution of proximity values, and (D) the product space matrix sorted in increasing order of the is numerical code.}
  \label{productSpace}
\end{figure}

Using this significant trade flow over 2000-2015, we can create $1242 \times 1242$ matrix whose entities are the proximity between products. 
Figure\ref{productSpace} shows the product space of world market in the period from 2000 to 2015. 

\clearpage

\section*{Appendix B. Regression Results}
\label{appendixB}

\begin{table}[!h] \centering 
  \caption{Bilateral trade volume after two years for periods 2000-2006} 
  \label{SI_main1} 
  \begin{adjustbox}{width=1\textwidth,center}
\begin{tabular}{@{\extracolsep{5pt}}lD{.}{.}{-3} D{.}{.}{-3} D{.}{.}{-3} D{.}{.}{-3} D{.}{.}{-3} D{.}{.}{-3} D{.}{.}{-3} } 
\\[-1.8ex]\hline 
\hline \\[-1.8ex] 
 & \multicolumn{7}{c}{\textit{Dependent variable}:~log $x^{t+2}_{opd}$} \\ 
\cline{2-8} 
\\[-1.8ex] & \multicolumn{1}{c}{(1)} & \multicolumn{1}{c}{(2)} & \multicolumn{1}{c}{(3)} & \multicolumn{1}{c}{(4)} & \multicolumn{1}{c}{(5)} & \multicolumn{1}{c}{(6)} & \multicolumn{1}{c}{(7)}\\ 
\hline \\[-1.8ex] 
 $\omega^t_{opd}$ &  & 0.122^{***} &  &  & 0.116^{***} & 0.202^{***} & 0.209^{***} \\ 
  &  & (0.001) &  &  & (0.001) & (0.001) & (0.001) \\ 
 $\Omega^{(d)}_{opd}$ &  &  & 0.118^{***} &  & 0.109^{***} & 0.139^{***} & 0.143^{***} \\ 
  &  &  & (0.001) &  & (0.001) & (0.001) & (0.001) \\ 
  $\Omega^{(o)}_{opd}$ &  &  &  & 0.041^{***} & 0.027^{***} & 0.042^{***} & 0.077^{***} \\ 
  &  &  &  & (0.001) & (0.001) & (0.001) & (0.001) \\ 
  log $x^t_{opd}$ & 0.441^{***} & 0.439^{***} & 0.441^{***} & 0.441^{***} & 0.440^{***} & 0.396^{***} & 0.377^{***} \\ 
  & (0.0003) & (0.0003) & (0.0003) & (0.0003) & (0.0003) & (0.0003) & (0.0003) \\ 
  log $x^t_{op}$ & 0.335^{***} & 0.326^{***} & 0.338^{***} & 0.333^{***} & 0.327^{***} & 0.318^{***} & 0.344^{***} \\ 
  & (0.0003) & (0.0003) & (0.0003) & (0.0003) & (0.0003) & (0.0004) & (0.0005) \\ 
  log $x^t_{pd}$ & 0.222^{***} & 0.225^{***} & 0.220^{***} & 0.222^{***} & 0.223^{***} & 0.201^{***} & 0.206^{***} \\ 
  & (0.0003) & (0.0003) & (0.0003) & (0.0003) & (0.0003) & (0.0005) & (0.0005) \\ 
  log $Distance$ & -1.073^{***} & -1.055^{***} & -0.941^{***} & -1.029^{***} & -0.903^{***} & -1.283^{***} & -1.046^{***} \\ 
  & (0.002) & (0.002) & (0.002) & (0.002) & (0.002) & (0.002) & (0.003) \\ 
  log $gdp^t_o$ &  &  &  &  &  & 0.247^{***} & 0.272^{***} \\ 
  &  &  &  &  &  & (0.002) & (0.002) \\ 
  log $gdp^t_d$ &  &  &  &  &  & 0.285^{***} & 0.334^{***} \\ 
  &  &  &  &  &  & (0.002) & (0.002) \\ 
  log $Population_o$ &  &  &  &  &  & 0.691^{***} & 0.680^{***} \\ 
  &  &  &  &  &  & (0.002) & (0.002) \\ 
  log $Population_d$  &  &  &  &  &  & 0.521^{***} & 0.525^{***} \\ 
  &  &  &  &  &  & (0.002) & (0.002) \\ 
  $Border_{od}$ &  &  &  &  &  &  & 0.712^{***} \\ 
  &  &  &  &  &  &  & (0.003) \\ 
  $Colony_{od}$  &  &  &  &  &  &  & 0.052^{***} \\ 
  &  &  &  &  &  &  & (0.003) \\ 
  $Language_{od}$ &  &  &  &  &  &  & 0.545^{***} \\ 
  &  &  &  &  &  &  & (0.002) \\ 
  log$Lang. Proximity_{od}$ &  &  &  &  &  &  & 0.031^{***} \\ 
  &  &  &  &  &  &  & (0.001) \\ 
  Constant & 0.407^{***} & 0.465^{***} & -0.080^{***} & 0.279^{***} & -0.077^{***} & -8.804^{***} & -10.354^{***} \\ 
  & (0.008) & (0.008) & (0.009) & (0.009) & (0.009) & (0.018) & (0.019) \\ 
 \hline \\[-1.8ex] 
Observations & \multicolumn{1}{c}{10,911,584} & \multicolumn{1}{c}{10,911,584} & \multicolumn{1}{c}{10,911,584} & \multicolumn{1}{c}{10,911,584} & \multicolumn{1}{c}{10,911,584} & \multicolumn{1}{c}{10,911,584} & \multicolumn{1}{c}{10,911,584} \\ 
Adjusted R$^{2}$ & \multicolumn{1}{c}{0.469} & \multicolumn{1}{c}{0.470} & \multicolumn{1}{c}{0.470} & \multicolumn{1}{c}{0.469} & \multicolumn{1}{c}{0.471} & \multicolumn{1}{c}{0.489} & \multicolumn{1}{c}{0.494} \\ 
Residual Std. Error & \multicolumn{1}{c}{2.632} & \multicolumn{1}{c}{2.629} & \multicolumn{1}{c}{2.630} & \multicolumn{1}{c}{2.632} & \multicolumn{1}{c}{2.628} & \multicolumn{1}{c}{2.583} & \multicolumn{1}{c}{2.568} \\ 
\hline 
\hline \\[-1.8ex] 
\textit{Note:}  & \multicolumn{7}{r}{$^{*}$p$<$0.1; $^{**}$p$<$0.05; $^{***}$p$<$0.01} \\ 
\end{tabular} 
\end{adjustbox}
\end{table}

\begin{table*}
  \caption{Bilateral trade volume after two years for periods 2000-2006 (pre-financial crisis), 2007-2012(crisis period) and 2012-2015 (recovery period)} 
  \label{SI_main2} 
    \begin{adjustbox}{width=1\textwidth,center}
\begin{tabular}{@{\extracolsep{5pt}}lD{.}{.}{-3} D{.}{.}{-3} D{.}{.}{-3} D{.}{.}{-3} D{.}{.}{-3} D{.}{.}{-3}} 
\\[-1.8ex]\hline 
\hline \\[-1.8ex] 
 & \multicolumn{6}{c}{\textit{Dependent variable}:~log $x^{t+2}_{opd}$} \\ 
\cline{2-7} 
  & \multicolumn{2}{c}{\textit{2000-2006}} & \multicolumn{2}{c}{\textit{2007-2012}}  & \multicolumn{2}{c}{\textit{2012-2015}} \\ 
  \cline{2-3} \cline{4-5} \cline{6-7} 
\\[-1.8ex] & \multicolumn{1}{c}{(1)} & \multicolumn{1}{c}{(2)}& \multicolumn{1}{c}{(3)}& \multicolumn{1}{c}{(4)}& \multicolumn{1}{c}{(5)}& \multicolumn{1}{c}{(6)}\\ 
\hline \\[-1.8ex] 
  $\omega^t_{opd}$ & 0.116^{***} & 0.209^{***} & 0.082^{***} & 0.180^{***} &0.054^{***} & 0.152^{***}\\ 
  & (0.001) & (0.001)  & (0.001) & (0.001)  & (0.001) & (0.001)\\ 
  $\Omega^{(d)}_{opd}$ & 0.109^{***} & 0.143^{***}& 0.105^{***} & 0.143^{***}  & 0.123^{***} & 0.152^{***} \\ 
  & (0.001) & (0.001)  & (0.001) & (0.001) & (0.001) & (0.001)\\ 
  $\Omega^{(o)}_{opd}$ & 0.027^{***} & 0.077^{***} & 0.073^{***} & 0.105^{***} & 0.087^{***} & 0.107^{***} \\ 
  & (0.001) & (0.001) & (0.001) & (0.001) & (0.001) & (0.001)\\ 
  log $x^t_{opd}$ & 0.440^{***} & 0.377^{***}& 0.485^{***} & 0.422^{***} & 0.526^{***} & 0.462^{***} \\ 
  & (0.0003) & (0.0003)& (0.0003) & (0.0003) & (0.0004) & (0.0004)\\ 
  log $x^t_{op}$ & 0.327^{***} & 0.344^{***}& 0.327^{***} & 0.342^{***} & 0.310^{***} & 0.325^{***} \\ 
  & (0.0003) & (0.0005) & (0.0004) & (0.001) & (0.0005) & (0.001) \\ 
  log $x^t_{pd}$ & 0.223^{***} & 0.206^{***} &0.178^{***} & 0.165^{***}  & 0.177^{***} & 0.159^{***} \\ 
  & (0.0003) & (0.0005) & (0.0004) & (0.001) & (0.0005) & (0.001)\\ 
  log $Distance$ & -0.903^{***}  & -1.046^{***}& -0.758^{***} & -0.919^{***} & -0.728^{***} & -0.954^{***}\\ 
  & (0.002)  & (0.003)& (0.003) & (0.003) & (0.003) & (0.004)\\ 
  log $gdp^t_o$ & & 0.272^{***} &  & 0.363^{***} &  & 0.406^{***} \\ 
  &  & (0.002) &  & (0.002) &  & (0.003) \\ 
  log $gdp^t_d$ &  & 0.334^{***}&  & 0.355^{***} &  & 0.480^{***}\\ 
  &  & (0.002) &  & (0.002) &  & (0.002)\\ 
  log $Population_o$ &  & 0.680^{***}&  & 0.654^{***} &  & 0.654^{***} \\ 
  &  & (0.002)&  & (0.002) &  & (0.002)\\ 
  log $Population_d$  &  & 0.525^{***}&  & 0.560^{***} &  & 0.518^{***}\\ 
   & & (0.002) &  & (0.002) &  & (0.002) \\ 
  $Border_{od}$ &  &  0.712^{***}&  & 0.717^{***} &  & 0.611^{***}\\ 
  &  & (0.003)&  & (0.004) &  & (0.005)\\ 
  $Colony_{od}$ &  & 0.052^{***} &  & 0.127^{***} &  & 0.193^{***}\\ 
   &  & (0.003)&  & (0.004) &  & (0.005) \\ 
  $Language_{od}$ &  & 0.545^{***} &  & 0.441^{***} &  & 0.334^{***}\\ 
  &  & (0.002) &  & (0.003) &  & (0.003) \\ 
  log$Lang. Proximity_{od}$  &  & 0.031^{***}&  & 0.027^{***} &  & -0.001 \\ 
  &  & (0.001)&  & (0.001) &  & (0.001) \\ 
  Constant & -0.077^{***} & -10.354^{***}    & -0.585^{***} & -11.482^{***} & -1.355^{***} & -12.762^{***}\\ 
  & (0.009) & (0.019)  & (0.012) & (0.025) & (0.015) & (0.030)  \\ 
 \hline \\[-1.8ex] 
Observations & \multicolumn{1}{c}{10,911,584} & \multicolumn{1}{c}{10,911,584}& \multicolumn{1}{c}{7,591,489} & \multicolumn{1}{c}{7,591,489} & \multicolumn{1}{c}{5,332,257} & \multicolumn{1}{c}{5,332,257}\\ 
Adjusted R$^{2}$ &\multicolumn{1}{c}{0.471} & \multicolumn{1}{c}{0.494} & \multicolumn{1}{c}{0.496} & \multicolumn{1}{c}{0.516} &  \multicolumn{1}{c}{0.539} & \multicolumn{1}{c}{0.558}\\ 
Residual Std. Error &\multicolumn{1}{c}{2.628} & \multicolumn{1}{c}{2.568}& \multicolumn{1}{c}{2.691} & \multicolumn{1}{c}{2.637}& \multicolumn{1}{c}{2.582} & \multicolumn{1}{c}{2.529} \\ 
\hline 
\hline \\[-1.8ex] 
\textit{Note:}  & \multicolumn{2}{r}{$^{*}$p$<$0.1; $^{**}$p$<$0.05; $^{***}$p$<$0.01} \\ 
\end{tabular} 
\end{adjustbox}
\end{table*} 

\clearpage
\section*{Appendix C. Summary statistics and correlation table}
\label{appendixC}

\begin{table}[!htbp] \centering 
  \caption{Summary statistics: 2000-2006} 
  \label{} 
\begin{tabular}{@{\extracolsep{5pt}}lccccc} 
\\[-1.8ex]\hline 
\hline \\[-1.8ex] 
Statistic & \multicolumn{1}{c}{N} & \multicolumn{1}{c}{Mean} & \multicolumn{1}{c}{St. Dev.} & \multicolumn{1}{c}{Min} & \multicolumn{1}{c}{Max} \\ 
\hline \\[-1.8ex] 
  $\omega^t_{opd}$ & 10,911,584 & 0.000 & 1.000 & $-$3.366 & 26.699 \\ 
 $\Omega^{(d)}_{opd}$ & 10,911,584 & 0.000 & 1.000 & $-$0.959 & 93.647 \\ 
$\Omega^{(o)}_{opd}$  & 10,911,584 & 0.000 & 1.000 & $-$1.172 & 51.786 \\ 
log $x^t_{opd}$ & 10,911,584 & 0.000 & 1.000 & $-$1.339 & 4.086 \\ 
log $x^t_{op}$ & 10,911,584 & 0.000 & 1.000 & $-$2.547 & 3.172 \\ 
log $x^t_{pd}$ & 10,911,584 & 0.000 & 1.000 & $-$2.544 & 3.758 \\ 
log $Distance$ & 10,911,584 & 0.000 & 1.000 & $-$3.855 & 1.578 \\ 
log $gdp^t_o$ & 10,911,584 & 0.000 & 1.000 & $-$3.118 & 1.339 \\ 
log $gdp^t_d$  & 10,911,584 & 0.000 & 1.000 & $-$2.477 & 1.541 \\ 
log $Population_o$ & 10,911,584 & 0.000 & 1.000 & $-$2.474 & 2.344 \\ 
log $Population_d$ & 10,911,584 & 0.000 & 1.000 & $-$2.324 & 2.768 \\ 
$Border_{od}$ & 10,911,584 & 0.089 & 0.284 & 0 & 1 \\ 
$Colony_{od}$ & 10,911,584 & 0.060 & 0.237 & 0 & 1 \\ 
$Language_{od}$ & 10,911,584 & 0.155 & 0.362 & 0 & 1 \\ 
log$Lang. Proximity_{od}$ & 10,911,584 & 0.000 & 1.000 & $-$0.312 & 5.050 \\ 
\hline \\[-1.8ex] 
\end{tabular} 
\end{table} 

\begin{table}[!htbp] \centering 
  \caption{Summary statistics: 2007-2012} 
  \label{} 
\begin{tabular}{@{\extracolsep{5pt}}lccccc} 
\\[-1.8ex]\hline 
\hline \\[-1.8ex] 
Statistic & \multicolumn{1}{c}{N} & \multicolumn{1}{c}{Mean} & \multicolumn{1}{c}{St. Dev.} & \multicolumn{1}{c}{Min} & \multicolumn{1}{c}{Max} \\ 
\hline \\[-1.8ex] 
$\omega^t_{opd}$ & 7,591,489 & 0.000 & 1.000 & $-$3.347 & 26.234 \\ 
$\Omega^{(d)}_{opd}$ & 7,591,489 & 0.000 & 1.000 & $-$0.915 & 91.287 \\ 
$\Omega^{(o)}_{opd}$  & 7,591,489 & 0.000 & 1.000 & $-$1.171 & 25.338 \\ 
log $x^t_{opd}$ & 7,591,489 & 0.000 & 1.000 & $-$1.374 & 3.977 \\ 
log $x^t_{op}$ & 7,591,489 & 0.000 & 1.000 & $-$2.673 & 3.094 \\ 
log $x^t_{pd}$ & 7,591,489 & 0.000 & 1.000 & $-$2.830 & 3.745 \\ 
log $Distance$ & 7,591,489 & 0.000 & 1.000 & $-$3.951 & 1.580 \\ 
log $gdp^t_o$ & 7,591,489 & 0.000 & 1.000 & $-$3.285 & 1.479 \\ 
log $gdp^t_d$  & 7,591,489 & 0.000 & 1.000 & $-$2.633 & 1.637 \\ 
log $Population_o$ & 7,591,489 & 0.000 & 1.000 & $-$1.990 & 2.328 \\ 
log $Population_d$ & 7,591,489 & 0.000 & 1.000 & $-$1.916 & 2.749 \\ 
$Border_{od}$ & 7,591,489 & 0.082 & 0.274 & 0 & 1 \\ 
$Colony_{od}$ & 7,591,489 & 0.053 & 0.224 & 0 & 1 \\ 
$Language_{od}$ & 7,591,489 & 0.153 & 0.360 & 0 & 1 \\ 
log$Lang. Proximity_{od}$  & 7,591,489 & $-$0.000 & 1.000 & $-$0.298 & 5.349 \\ 
\hline \\[-1.8ex] 
\end{tabular} 
\end{table}

\begin{table}[!htbp] \centering 
  \caption{Summary statistics: 2012-2015} 
  \label{} 
\begin{tabular}{@{\extracolsep{5pt}}lccccc} 
\\[-1.8ex]\hline 
\hline \\[-1.8ex] 
Statistic & \multicolumn{1}{c}{N} & \multicolumn{1}{c}{Mean} & \multicolumn{1}{c}{St. Dev.} & \multicolumn{1}{c}{Min} & \multicolumn{1}{c}{Max} \\ 
\hline \\[-1.8ex] 
  $\omega^t_{opd}$ & 5,332,257 & 0.000 & 1.000 & $-$3.373 & 24.428 \\ 
 $\Omega^{(d)}_{opd}$ & 5,332,257 & 0.000 & 1.000 & $-$0.965 & 98.891 \\ 
$\Omega^{(o)}_{opd}$  & 5,332,257 & 0.000 & 1.000 & $-$1.185 & 34.256 \\ 
log $x^t_{opd}$ & 5,332,257 & 0.000 & 1.000 & $-$1.377 & 3.976 \\ 
log $x^t_{op}$ & 5,332,257 & 0.000 & 1.000 & $-$2.787 & 3.116 \\ 
log $x^t_{pd}$ & 5,332,257 & 0.000 & 1.000 & $-$2.900 & 3.591 \\ 
log $Distance$ & 5,332,257 & 0.000 & 1.000 & $-$3.993 & 1.581 \\ 
log $gdp^t_o$ & 5,332,257 & 0.000 & 1.000 & $-$3.193 & 1.527 \\ 
log $gdp^t_d$  & 5,332,257 & 0.000 & 1.000 & $-$2.393 & 1.652 \\ 
log $Population_o$ & 5,332,257 & 0.000 & 1.000 & $-$1.994 & 2.340 \\ 
log $Population_d$ & 5,332,257 & 0.000 & 1.000 & $-$1.861 & 2.754 \\ 
$Border_{od}$ & 5,332,257 & 0.078 & 0.268 & 0 & 1 \\ 
$Colony_{od}$ & 5,332,257 & 0.052 & 0.222 & 0 & 1 \\ 
$Language_{od}$ & 5,332,257 & 0.141 & 0.348 & 0 & 1 \\ 
log$Lang. Proximity_{od}$  & 5,332,257 & 0.000 & 1.000 & $-$0.298 & 5.383 \\ 
\hline \\[-1.8ex] 
\end{tabular} 
\end{table} 

\newgeometry{margin=1cm} 
\begin{landscape}

\begin{table}[!htbp] \centering 
  \caption{Correlation Matrix: 2000-2006} 
  \label{} 
  \begin{adjustbox}{width=1.2\textwidth,center}
\begin{tabular}{@{\extracolsep{5pt}} cccccccccccccccc} 
\\[-1.8ex]\hline 
\hline \\[-1.8ex] 
 &   $\omega^t_{opd}$ &  $\Omega^{(d)}_{opd}$ & $\Omega^{(o)}_{opd}$  & log $x^t_{opd}$ & log $x^t_{op}$ & log $x^t_{pd}$ & log $Distance$ & log $gdp^t_o$ & log $gdp^t_d$  & log $Population_o$ & log $Population_d$ & $Border_{od}$ & $Colony_{od}$ & $Language_{od}$ & log$Lang. Proximity_{od}$  \\ 
\hline \\[-1.8ex] 
  $\omega^t_{opd}$ & $1$ & $0.038$ & $0.080$ & $0.132$ & $0.215$ & $ $-$0.020$ & $ $-$0.026$ & $ $-$0.006$ & $ $-$0.055$ & $ $-$0.060$ & $ $-$0.081$ & $ $-$0.017$ & $ $-$0.025$ & $ $-$0.035$ & $ $-$0.018$ \\ 
 $\Omega^{(d)}_{opd}$ & $0.038$ & $1$ & $0.287$ & $0.049$ & $ $-$0.182$ & $ $-$0.009$ & $ $-$0.536$ & $ $-$0.037$ & $0.091$ & $ $-$0.188$ & $ $-$0.150$ & $0.229$ & $0.013$ & $0.044$ & $0.095$ \\ 
$\Omega^{(o)}_{opd}$  & $0.080$ & $0.287$ & $1$ & $0.064$ & $ $-$0.003$ & $ $-$0.035$ & $ $-$0.457$ & $0.123$ & $0.025$ & $ $-$0.176$ & $ $-$0.123$ & $0.126$ & $0.005$ & $ $-$0.068$ & $0.100$ \\ 
log $x^t_{opd}$ & $0.132$ & $0.049$ & $0.064$ & $1$ & $0.459$ & $0.344$ & $ $-$0.117$ & $0.170$ & $0.159$ & $0.135$ & $0.104$ & $0.110$ & $0.063$ & $0.019$ & $0.132$ \\ 
log $x^t_{op}$ & $0.215$ & $ $-$0.182$ & $ $-$0.003$ & $0.459$ & $1$ & $0.171$ & $0.239$ & $0.345$ & $ $-$0.120$ & $0.288$ & $ $-$0.083$ & $ $-$0.190$ & $ $-$0.028$ & $ $-$0.163$ & $0.015$ \\ 
log $x^t_{pd}$ & $ $-$0.020$ & $ $-$0.009$ & $ $-$0.035$ & $0.344$ & $0.171$ & $1$ & $0.078$ & $ $-$0.038$ & $0.486$ & $ $-$0.140$ & $0.331$ & $ $-$0.072$ & $ $-$0.026$ & $ $-$0.083$ & $0.085$ \\ 
log $Distance$ & $ $-$0.026$ & $ $-$0.536$ & $ $-$0.457$ & $ $-$0.117$ & $0.239$ & $0.078$ & $1$ & $ $-$0.015$ & $ $-$0.066$ & $0.285$ & $0.211$ & $ $-$0.486$ & $ $-$0.042$ & $ $-$0.044$ & $ $-$0.217$ \\ 
log $gdp^t_o$ & $ $-$0.006$ & $ $-$0.037$ & $0.123$ & $0.170$ & $0.345$ & $ $-$0.038$ & $ $-$0.015$ & $1$ & $0.016$ & $ $-$0.338$ & $ $-$0.041$ & $ $-$0.115$ & $0.029$ & $ $-$0.072$ & $0.110$ \\ 
log $gdp^t_d$ & $ $-$0.055$ & $0.091$ & $0.025$ & $0.159$ & $ $-$0.120$ & $0.486$ & $ $-$0.066$ & $0.016$ & $1$ & $ $-$0.084$ & $ $-$0.148$ & $ $-$0.032$ & $ $-$0.001$ & $ $-$0.065$ & $0.138$ \\ 
log $Population_o$ & $ $-$0.060$ & $ $-$0.188$ & $ $-$0.176$ & $0.135$ & $0.288$ & $ $-$0.140$ & $0.285$ & $ $-$0.338$ & $ $-$0.084$ & $1$ & $ $-$0.013$ & $ $-$0.052$ & $0.027$ & $ $-$0.044$ & $0.004$ \\ 
log $Population_d$ & $ $-$0.081$ & $ $-$0.150$ & $ $-$0.123$ & $0.104$ & $ $-$0.083$ & $0.331$ & $0.211$ & $ $-$0.041$ & $ $-$0.148$ & $ $-$0.013$ & $1$ & $ $-$0.022$ & $0.019$ & $ $-$0.022$ & $0.058$ \\ 
$Border_{od}$ & $ $-$0.017$ & $0.229$ & $0.126$ & $0.110$ & $ $-$0.190$ & $ $-$0.072$ & $ $-$0.486$ & $ $-$0.115$ & $ $-$0.032$ & $ $-$0.052$ & $ $-$0.022$ & $1$ & $0.130$ & $0.170$ & $0.114$ \\ 
$Colony_{od}$ & $ $-$0.025$ & $0.013$ & $0.005$ & $0.063$ & $ $-$0.028$ & $ $-$0.026$ & $ $-$0.042$ & $0.029$ & $ $-$0.001$ & $0.027$ & $0.019$ & $0.130$ & $1$ & $0.264$ & $0.071$ \\ 
$Language_{od}$ & $ $-$0.035$ & $0.044$ & $ $-$0.068$ & $0.019$ & $ $-$0.163$ & $ $-$0.083$ & $ $-$0.044$ & $ $-$0.072$ & $ $-$0.065$ & $ $-$0.044$ & $ $-$0.022$ & $0.170$ & $0.264$ & $1$ & $ $-$0.103$ \\ 
log$Lang. Proximity_{od}$ & $ $-$0.018$ & $0.095$ & $0.100$ & $0.132$ & $0.015$ & $0.085$ & $ $-$0.217$ & $0.110$ & $0.138$ & $0.004$ & $0.058$ & $0.114$ & $0.071$ & $ $-$0.103$ & $1$ \\ 
\hline \\[-1.8ex] 
\end{tabular} 
\end{adjustbox}
\end{table} 

\begin{table}[!htbp] \centering 
  \caption{Correlation Matrix: 2007-2012} 
  \label{} 
    \begin{adjustbox}{width=1.2\textwidth,center}
\begin{tabular}{@{\extracolsep{5pt}} cccccccccccccccc} 
\\[-1.8ex]\hline 
\hline \\[-1.8ex] 
 &   $\omega^t_{opd}$ &  $\Omega^{(d)}_{opd}$ & $\Omega^{(o)}_{opd}$  & log $x^t_{opd}$ & log $x^t_{op}$ & log $x^t_{pd}$ & log $Distance$ & log $gdp^t_o$ & log $gdp^t_d$  & log $Population_o$ & log $Population_d$ & $Border_{od}$ & $Colony_{od}$ & $Language_{od}$ & log$Lang. Proximity_{od}$  \\  
\hline \\[-1.8ex] 
$\omega^t_{opd}$ & $1$ & $0.036$ & $0.083$ & $0.139$ & $0.233$ & $$-$0.012$ & $$-$0.030$ & $$-$0.001$ & $$-$0.049$ & $$-$0.066$ & $$-$0.086$ & $$-$0.018$ & $$-$0.024$ & $$-$0.027$ & $$-$0.015$ \\ 
$\Omega^{(d)}_{opd}$ & $0.036$ & $1$ & $0.273$ & $0.059$ & $$-$0.161$ & $$-$0.022$ & $$-$0.512$ & $$-$0.024$ & $0.093$ & $$-$0.179$ & $$-$0.153$ & $0.217$ & $0.013$ & $0.053$ & $0.087$ \\ 
$\Omega^{(o)}_{opd}$ & $0.083$ & $0.273$ & $1$ & $0.086$ & $$-$0.004$ & $$-$0.036$ & $$-$0.452$ & $0.139$ & $0.061$ & $$-$0.167$ & $$-$0.135$ & $0.129$ & $0.004$ & $$-$0.050$ & $0.101$ \\ 
log $x^t_{opd}$ & $0.139$ & $0.059$ & $0.086$ & $1$ & $0.481$ & $0.343$ & $$-$0.128$ & $0.137$ & $0.138$ & $0.166$ & $0.100$ & $0.126$ & $0.057$ & $0.004$ & $0.132$ \\ 
log $x^t_{op}$ & $0.233$ & $$-$0.161$ & $$-$0.004$ & $0.481$ & $1$ & $0.206$ & $0.219$ & $0.306$ & $$-$0.107$ & $0.298$ & $$-$0.088$ & $$-$0.168$ & $$-$0.028$ & $$-$0.182$ & $0.021$ \\ 
log $x^t_{pd}$ & $$-$0.012$ & $$-$0.022$ & $$-$0.036$ & $0.343$ & $0.206$ & $1$ & $0.078$ & $$-$0.036$ & $0.414$ & $$-$0.149$ & $0.342$ & $$-$0.063$ & $$-$0.017$ & $$-$0.091$ & $0.083$ \\ 
log $Distance$  & $$-$0.030$ & $$-$0.512$ & $$-$0.452$ & $$-$0.128$ & $0.219$ & $0.078$ & $1$ & $$-$0.043$ & $$-$0.104$ & $0.290$ & $0.224$ & $$-$0.477$ & $$-$0.046$ & $$-$0.053$ & $$-$0.213$ \\ 
log $gdp^t_o$ & $$-$0.001$ & $$-$0.024$ & $0.139$ & $0.137$ & $0.306$ & $$-$0.036$ & $$-$0.043$ & $1$ & $0.034$ & $$-$0.349$ & $$-$0.053$ & $$-$0.090$ & $0.038$ & $$-$0.117$ & $0.131$ \\ 
log $gdp^t_d$ & $$-$0.049$ & $0.093$ & $0.061$ & $0.138$ & $$-$0.107$ & $0.414$ & $$-$0.104$ & $0.034$ & $1$ & $$-$0.097$ & $$-$0.177$ & $$-$0.014$ & $0.012$ & $$-$0.096$ & $0.160$ \\ 
log $Population_o$  & $$-$0.066$ & $$-$0.179$ & $$-$0.167$ & $0.166$ & $0.298$ & $$-$0.149$ & $0.290$ & $$-$0.349$ & $$-$0.097$ & $1$ & $$-$0.012$ & $$-$0.046$ & $0.025$ & $$-$0.034$ & $0.001$ \\ 
log $Population_d$ & $$-$0.086$ & $$-$0.153$ & $$-$0.135$ & $0.100$ & $$-$0.088$ & $0.342$ & $0.224$ & $$-$0.053$ & $$-$0.177$ & $$-$0.012$ & $1$ & $$-$0.020$ & $0.022$ & $$-$0.007$ & $0.045$ \\ 
$Border_{od}$ & $$-$0.018$ & $0.217$ & $0.129$ & $0.126$ & $$-$0.168$ & $$-$0.063$ & $$-$0.477$ & $$-$0.090$ & $$-$0.014$ & $$-$0.046$ & $$-$0.020$ & $1$ & $0.126$ & $0.169$ & $0.111$ \\ 
$Colony_{od}$ & $$-$0.024$ & $0.013$ & $0.004$ & $0.057$ & $$-$0.028$ & $$-$0.017$ & $$-$0.046$ & $0.038$ & $0.012$ & $0.025$ & $0.022$ & $0.126$ & $1$ & $0.252$ & $0.070$ \\ 
$Language_{od}$ & $$-$0.027$ & $0.053$ & $$-$0.050$ & $0.004$ & $$-$0.182$ & $$-$0.091$ & $$-$0.053$ & $$-$0.117$ & $$-$0.096$ & $$-$0.034$ & $$-$0.007$ & $0.169$ & $0.252$ & $1$ & $$-$0.099$ \\ 
log$Lang. Proximity_{od}$ & $$-$0.015$ & $0.087$ & $0.101$ & $0.132$ & $0.021$ & $0.083$ & $$-$0.213$ & $0.131$ & $0.160$ & $0.001$ & $0.045$ & $0.111$ & $0.070$ & $$-$0.099$ & $1$ \\ 
\hline \\[-1.8ex] 
\end{tabular} 
\end{adjustbox}
\end{table} 

\begin{table}[!htbp] \centering 
  \caption{Correlation Matrix: 2012-2015} 
  \label{} 
    \begin{adjustbox}{width=1.2\textwidth,center}
\begin{tabular}{@{\extracolsep{5pt}} cccccccccccccccc} 
\\[-1.8ex]\hline 
\hline \\[-1.8ex] 
 &   $\omega^t_{opd}$ &  $\Omega^{(d)}_{opd}$ & $\Omega^{(o)}_{opd}$  & log $x^t_{opd}$ & log $x^t_{op}$ & log $x^t_{pd}$ & log $Distance$ & log $gdp^t_o$ & log $gdp^t_d$  & log $Population_o$ & log $Population_d$ & $Border_{od}$ & $Colony_{od}$ & $Language_{od}$ & log$Lang. Proximity_{od}$  \\ 
\hline \\[-1.8ex] 
$\omega^t_{opd}$ & $1$ & $0.033$ & $0.084$ & $0.143$ & $0.235$ & $$-$0.006$ & $$-$0.025$ & $$-$0.012$ & $$-$0.044$ & $$-$0.064$ & $$-$0.084$ & $$-$0.021$ & $$-$0.023$ & $$-$0.026$ & $$-$0.014$ \\ 
$\Omega^{(d)}_{opd}$ & $0.033$ & $1$ & $0.283$ & $0.069$ & $$-$0.167$ & $$-$0.023$ & $$-$0.542$ & $$-$0.036$ & $0.095$ & $$-$0.180$ & $$-$0.154$ & $0.236$ & $0.017$ & $0.044$ & $0.095$ \\ 
$\Omega^{(o)}_{opd}$ & $0.084$ & $0.283$ & $1$ & $0.082$ & $$-$0.015$ & $$-$0.042$ & $$-$0.440$ & $0.118$ & $0.059$ & $$-$0.161$ & $$-$0.132$ & $0.125$ & $0.002$ & $$-$0.053$ & $0.101$ \\ 
log $x^t_{opd}$  & $0.143$ & $0.069$ & $0.082$ & $1$ & $0.480$ & $0.350$ & $$-$0.125$ & $0.109$ & $0.128$ & $0.185$ & $0.108$ & $0.138$ & $0.061$ & $0.022$ & $0.124$ \\ 
log $x^t_{op}$ & $0.235$ & $$-$0.167$ & $$-$0.015$ & $0.480$ & $1$ & $0.213$ & $0.222$ & $0.271$ & $$-$0.124$ & $0.309$ & $$-$0.083$ & $$-$0.157$ & $$-$0.031$ & $$-$0.162$ & $0.009$ \\ 
log $x^t_{pd}$ & $$-$0.006$ & $$-$0.023$ & $$-$0.042$ & $0.350$ & $0.213$ & $1$ & $0.085$ & $$-$0.055$ & $0.387$ & $$-$0.148$ & $0.349$ & $$-$0.054$ & $$-$0.021$ & $$-$0.073$ & $0.076$ \\ 
log $Distance$  & $$-$0.025$ & $$-$0.542$ & $$-$0.440$ & $$-$0.125$ & $0.222$ & $0.085$ & $1$ & $$-$0.010$ & $$-$0.093$ & $0.283$ & $0.227$ & $$-$0.472$ & $$-$0.051$ & $$-$0.042$ & $$-$0.215$ \\ 
log $gdp^t_o$ & $$-$0.012$ & $$-$0.036$ & $0.118$ & $0.109$ & $0.271$ & $$-$0.055$ & $$-$0.010$ & $1$ & $0.019$ & $$-$0.332$ & $$-$0.048$ & $$-$0.089$ & $0.030$ & $$-$0.104$ & $0.116$ \\ 
log $gdp^t_d$ & $$-$0.044$ & $0.095$ & $0.059$ & $0.128$ & $$-$0.124$ & $0.387$ & $$-$0.093$ & $0.019$ & $1$ & $$-$0.101$ & $$-$0.169$ & $$-$0.006$ & $0.008$ & $$-$0.075$ & $0.158$ \\ 
log $Population_o$  & $$-$0.064$ & $$-$0.180$ & $$-$0.161$ & $0.185$ & $0.309$ & $$-$0.148$ & $0.283$ & $$-$0.332$ & $$-$0.101$ & $1$ & $$-$0.007$ & $$-$0.035$ & $0.028$ & $$-$0.005$ & $$-$0.001$ \\ 
log $Population_d$ & $$-$0.084$ & $$-$0.154$ & $$-$0.132$ & $0.108$ & $$-$0.083$ & $0.349$ & $0.227$ & $$-$0.048$ & $$-$0.169$ & $$-$0.007$ & $1$ & $$-$0.020$ & $0.021$ & $0.003$ & $0.042$ \\ 
$Border_{od}$ & $$-$0.021$ & $0.236$ & $0.125$ & $0.138$ & $$-$0.157$ & $$-$0.054$ & $$-$0.472$ & $$-$0.089$ & $$-$0.006$ & $$-$0.035$ & $$-$0.020$ & $1$ & $0.136$ & $0.165$ & $0.114$ \\ 
$Colony_{od}$ & $$-$0.023$ & $0.017$ & $0.002$ & $0.061$ & $$-$0.031$ & $$-$0.021$ & $$-$0.051$ & $0.030$ & $0.008$ & $0.028$ & $0.021$ & $0.136$ & $1$ & $0.260$ & $0.070$ \\ 
$Language_{od}$ & $$-$0.026$ & $0.044$ & $$-$0.053$ & $0.022$ & $$-$0.162$ & $$-$0.073$ & $$-$0.042$ & $$-$0.104$ & $$-$0.075$ & $$-$0.005$ & $0.003$ & $0.165$ & $0.260$ & $1$ & $$-$0.092$ \\ 
log$Lang. Proximity_{od}$  & $$-$0.014$ & $0.095$ & $0.101$ & $0.124$ & $0.009$ & $0.076$ & $$-$0.215$ & $0.116$ & $0.158$ & $$-$0.001$ & $0.042$ & $0.114$ & $0.070$ & $$-$0.092$ & $1$ \\ 
\hline \\[-1.8ex] 
\end{tabular} 
\end{adjustbox}
\end{table} 
\end{landscape}
\restoregeometry

\clearpage

\section*{Appendix D. Relationship between bilateral trade volume after two years and the three learning channels by products' competitiveness}
\label{appendixD}

\begin{table}[!htbp] \centering 
  \caption{Summary statistics: New exporters} 
  \label{} 
\begin{tabular}{@{\extracolsep{5pt}}lccccc} 
\\[-1.8ex]\hline 
\hline \\[-1.8ex] 
Statistic & \multicolumn{1}{c}{N} & \multicolumn{1}{c}{Mean} & \multicolumn{1}{c}{St. Dev.} & \multicolumn{1}{c}{Min} & \multicolumn{1}{c}{Max} \\ 
\hline \\[-1.8ex] 
$\omega^t_{opd}$ & 922,092 & 0.000 & 1.000 & $-$2.520 & 15.486 \\ 
$\Omega^{(d)}_{opd}$  & 922,092 & 0.000 & 1.000 & $-$0.720 & 55.607 \\ 
$\Omega^{(o)}_{opd}$ & 922,092 & 0.000 & 1.000 & $-$1.167 & 25.297 \\ 
log $x^t_{opd}$  & 922,092 & 0.000 & 1.000 & $-$1.000 & 4.508 \\ 
log $x^t_{op}$ & 922,092 & 0.000 & 1.000 & $-$1.415 & 3.296 \\ 
log $x^t_{pd}$ & 922,092 & 0.000 & 1.000 & $-$2.918 & 2.886 \\ 
log $Distance$ & 922,092 & 0.000 & 1.000 & $-$3.309 & 1.785 \\ 
log $gdp^t_o$ & 922,092 & 0.000 & 1.000 & $-$2.429 & 2.330 \\ 
log $gdp^t_d$ & 922,092 & 0.000 & 1.000 & $-$2.477 & 1.455 \\ 
log $Population_o$ & 922,092 & 0.000 & 1.000 & $-$2.220 & 3.552 \\ 
log $Population_d$ & 922,092 & 0.000 & 1.000 & $-$2.383 & 2.676 \\ 
$Border_{od}$ & 922,092 & 0.170 & 0.376 & 0 & 1 \\ 
$Colony_{od}$ & 922,092 & 0.058 & 0.234 & 0 & 1 \\ 
$Language_{od}$ & 922,092 & 0.292 & 0.455 & 0 & 1 \\ 
log$Lang. Proximity_{od}$  & 922,092 & 0.000 & 1.000 & $-$0.239 & 7.600 \\ 
\hline \\[-1.8ex] 
\end{tabular} 
\end{table} 

\begin{table}[!htbp] \centering 
  \caption{Summary statistics: Nascent exporters} 
  \label{} 
\begin{tabular}{@{\extracolsep{5pt}}lccccc} 
\\[-1.8ex]\hline 
\hline \\[-1.8ex] 
Statistic & \multicolumn{1}{c}{N} & \multicolumn{1}{c}{Mean} & \multicolumn{1}{c}{St. Dev.} & \multicolumn{1}{c}{Min} & \multicolumn{1}{c}{Max} \\ 
\hline \\[-1.8ex] 
$\omega^t_{opd}$ & 463,388 & 0.000 & 1.000 & $-$3.112 & 15.288 \\ 
$\Omega^{(d)}_{opd}$ & 463,388 & 0.000 & 1.000 & $-$1.050 & 91.632 \\ 
$\Omega^{(o)}_{opd}$  & 463,388 & 0.000 & 1.000 & $-$1.248 & 25.615 \\ 
log $x^t_{opd}$ & 463,388 & 0.000 & 1.000 & $-$1.256 & 4.077 \\ 
log $x^t_{op}$ & 463,388 & 0.000 & 1.000 & $-$2.905 & 3.434 \\ 
log $x^t_{pd}$ & 463,388 & 0.000 & 1.000 & $-$3.336 & 3.460 \\ 
log $Distance$ & 463,388 & 0.000 & 1.000 & $-$3.659 & 1.581 \\ 
log $gdp^t_o$  & 463,388 & 0.000 & 1.000 & $-$3.222 & 1.739 \\ 
log $gdp^t_d$  & 463,388 & 0.000 & 1.000 & $-$2.493 & 1.514 \\ 
log $Population_o$ & 463,388 & 0.000 & 1.000 & $-$2.137 & 2.852 \\ 
log $Population_d$ & 463,388 & 0.000 & 1.000 & $-$2.327 & 2.811 \\ 
$Border_{od}$ & 463,388 & 0.091 & 0.287 & 0 & 1 \\ 
$Colony_{od}$  & 463,388 & 0.042 & 0.200 & 0 & 1 \\ 
$Language_{od}$ & 463,388 & 0.160 & 0.366 & 0 & 1 \\ 
log$Lang. Proximity_{od}$ & 463,388 & 0.000 & 1.000 & $-$0.299 & 5.712 \\ 
\hline \\[-1.8ex] 
\end{tabular} 
\end{table} 

\begin{table}[!htbp] \centering 
  \caption{Summary statistics: Experienced exporters} 
  \label{} 
\begin{tabular}{@{\extracolsep{5pt}}lccccc} 
\\[-1.8ex]\hline 
\hline \\[-1.8ex] 
Statistic & \multicolumn{1}{c}{N} & \multicolumn{1}{c}{Mean} & \multicolumn{1}{c}{St. Dev.} & \multicolumn{1}{c}{Min} & \multicolumn{1}{c}{Max} \\ 
\hline \\[-1.8ex] 
$\omega^t_{opd}$ & 8,045,262 & 0.000 & 1.000 & $-$3.738 & 28.851 \\ 
$\Omega^{(d)}_{opd}$ & 8,045,262 & 0.000 & 1.000 & $-$1.106 & 108.927 \\ 
$\Omega^{(o)}_{opd}$ & 8,045,262 & 0.000 & 1.000 & $-$1.178 & 50.902 \\ 
log $x^t_{opd}$ & 8,045,262 & 0.000 & 1.000 & $-$1.467 & 3.950 \\ 
log $x^t_{op}$ & 8,045,262 & 0.000 & 1.000 & $-$3.373 & 3.456 \\ 
log $x^t_{pd}$ & 8,045,262 & 0.000 & 1.000 & $-$2.559 & 4.114 \\ 
log $Distance$  & 8,045,262 & 0.000 & 1.000 & $-$4.085 & 1.552 \\ 
log $gdp^t_o$ & 8,045,262 & 0.000 & 1.000 & $-$3.623 & 1.234 \\ 
log $gdp^t_d$  & 8,045,262 & 0.000 & 1.000 & $-$2.472 & 1.570 \\ 
log $Population_o$  & 8,045,262 & 0.000 & 1.000 & $-$2.612 & 2.237 \\ 
log $Population_d$ & 8,045,262 & 0.000 & 1.000 & $-$2.316 & 2.801 \\ 
$Border_{od}$ & 8,045,262 & 0.069 & 0.254 & 0 & 1 \\ 
$Colony_{od}$ & 8,045,262 & 0.062 & 0.241 & 0 & 1 \\ 
$Language_{od}$ & 8,045,262 & 0.128 & 0.334 & 0 & 1 \\ 
log$Lang. Proximity_{od}$ & 8,045,262 & 0.000 & 1.000 & $-$0.325 & 4.763 \\ 
\hline \\[-1.8ex] 
\end{tabular} 
\end{table} 

\newgeometry{margin=1cm} 
\begin{landscape}

\begin{table}[!htbp] \centering 
  \caption{Correlation Matrix: New exporters} 
  \label{} 
      \begin{adjustbox}{width=1.2\textwidth,center}
\begin{tabular}{@{\extracolsep{5pt}} cccccccccccccccc} 
\\[-1.8ex]\hline 
\hline \\[-1.8ex] 
 & $\omega^t_{opd}$  & $\Omega^{(d)}_{opd}$ & $\Omega^{(o)}_{opd}$ & log $x^t_{opd}$ & log $x^t_{op}$ & log $x^t_{pd}$ & log $Distance$ & log $gdp^t_o$ & log $gdp^t_d$ & log $Population_o$ & log $Population_d$ & $Border_{od}$ & $Colony_{od}$ & $Language_{od}$ & log$Lang. Proximity_{od}$ \\ 
\hline \\[-1.8ex] 
$\omega^t_{opd}$  & $1$ & $0.064$ & $0.127$ & $0.071$ & $0.112$ & $$-$0.095$ & $$-$0.108$ & $0.065$ & $$-$0.107$ & $$-$0.125$ & $$-$0.117$ & $0.027$ & $$-$0.033$ & $0.035$ & $$-$0.033$ \\ 
$\Omega^{(d)}_{opd}$ & $0.064$ & $1$ & $0.241$ & $$-$0.001$ & $$-$0.077$ & $$-$0.166$ & $$-$0.387$ & $0.001$ & $$-$0.076$ & $$-$0.109$ & $$-$0.200$ & $0.165$ & $$-$0.031$ & $0.123$ & $0.015$ \\ 
$\Omega^{(o)}_{opd}$  & $0.127$ & $0.241$ & $1$ & $0.083$ & $$-$0.006$ & $$-$0.179$ & $$-$0.462$ & $0.083$ & $$-$0.112$ & $$-$0.151$ & $$-$0.189$ & $0.206$ & $0.014$ & $$-$0.014$ & $0.073$ \\ 
log $x^t_{opd}$ & $0.071$ & $$-$0.001$ & $0.083$ & $1$ & $0.429$ & $0.211$ & $$-$0.129$ & $0.154$ & $0.019$ & $0.027$ & $0.015$ & $0.114$ & $0.049$ & $0.043$ & $0.069$ \\ 
log $x^t_{op}$ & $0.112$ & $$-$0.077$ & $$-$0.006$ & $0.429$ & $1$ & $0.344$ & $0.200$ & $0.373$ & $$-$0.051$ & $0.113$ & $$-$0.074$ & $$-$0.193$ & $$-$0.088$ & $$-$0.198$ & $0.024$ \\ 
log $x^t_{pd}$ & $$-$0.095$ & $$-$0.166$ & $$-$0.179$ & $0.211$ & $0.344$ & $1$ & $0.395$ & $0.068$ & $0.611$ & $0.006$ & $0.438$ & $$-$0.322$ & $0.083$ & $$-$0.292$ & $0.079$ \\ 
log $Distance$ & $$-$0.108$ & $$-$0.387$ & $$-$0.462$ & $$-$0.129$ & $0.200$ & $0.395$ & $1$ & $$-$0.023$ & $0.224$ & $0.250$ & $0.335$ & $$-$0.563$ & $$-$0.002$ & $$-$0.212$ & $$-$0.106$ \\ 
log $gdp^t_o$ & $0.065$ & $0.001$ & $0.083$ & $0.154$ & $0.373$ & $0.068$ & $$-$0.023$ & $1$ & $0.064$ & $$-$0.442$ & $$-$0.083$ & $$-$0.098$ & $$-$0.065$ & $$-$0.080$ & $0.109$ \\ 
log $gdp^t_d$ & $$-$0.107$ & $$-$0.076$ & $$-$0.112$ & $0.019$ & $$-$0.051$ & $0.611$ & $0.224$ & $0.064$ & $1$ & $$-$0.057$ & $0.013$ & $$-$0.278$ & $0.086$ & $$-$0.202$ & $0.079$ \\ 
log $Population_o$ & $$-$0.125$ & $$-$0.109$ & $$-$0.151$ & $0.027$ & $0.113$ & $0.006$ & $0.250$ & $$-$0.442$ & $$-$0.057$ & $1$ & $0.041$ & $$-$0.031$ & $0.010$ & $$-$0.007$ & $0.006$ \\ 
log $Population_d$ & $$-$0.117$ & $$-$0.200$ & $$-$0.189$ & $0.015$ & $$-$0.074$ & $0.438$ & $0.335$ & $$-$0.083$ & $0.013$ & $0.041$ & $1$ & $$-$0.093$ & $0.135$ & $$-$0.158$ & $0.067$ \\ 
$Border_{od}$  & $0.027$ & $0.165$ & $0.206$ & $0.114$ & $$-$0.193$ & $$-$0.322$ & $$-$0.563$ & $$-$0.098$ & $$-$0.278$ & $$-$0.031$ & $$-$0.093$ & $1$ & $0.084$ & $0.229$ & $0.061$ \\ 
$Colony_{od}$ & $$-$0.033$ & $$-$0.031$ & $0.014$ & $0.049$ & $$-$0.088$ & $0.083$ & $$-$0.002$ & $$-$0.065$ & $0.086$ & $0.010$ & $0.135$ & $0.084$ & $1$ & $0.131$ & $0.167$ \\ 
$Language_{od}$ & $0.035$ & $0.123$ & $$-$0.014$ & $0.043$ & $$-$0.198$ & $$-$0.292$ & $$-$0.212$ & $$-$0.080$ & $$-$0.202$ & $$-$0.007$ & $$-$0.158$ & $0.229$ & $0.131$ & $1$ & $$-$0.113$ \\ 
log$Lang. Proximity_{od}$ & $$-$0.033$ & $0.015$ & $0.073$ & $0.069$ & $0.024$ & $0.079$ & $$-$0.106$ & $0.109$ & $0.079$ & $0.006$ & $0.067$ & $0.061$ & $0.167$ & $$-$0.113$ & $1$ \\ 
\hline \\[-1.8ex] 
\end{tabular} 
\end{adjustbox}
\end{table} 

\begin{table}[!htbp] \centering 
  \caption{Correlation Matrix: Nascent exporters} 
  \label{} 
      \begin{adjustbox}{width=1.2\textwidth,center}
\begin{tabular}{@{\extracolsep{5pt}} cccccccccccccccc} 
\\[-1.8ex]\hline 
\hline \\[-1.8ex] 
 & $\omega^t_{opd}$  & $\Omega^{(d)}_{opd}$ & $\Omega^{(o)}_{opd}$ & log $x^t_{opd}$ & log $x^t_{op}$ & log $x^t_{pd}$ & log $Distance$ & log $gdp^t_o$ & log $gdp^t_d$ & log $Population_o$ & log $Population_d$ & $Border_{od}$ & $Colony_{od}$ & $Language_{od}$ & log$Lang. Proximity_{od}$ \\ 
\hline \\[-1.8ex] 
$\omega^t_{opd}$  & $1$ & $0.102$ & $0.122$ & $0.071$ & $0.114$ & $$-$0.018$ & $$-$0.135$ & $$-$0.039$ & $$-$0.070$ & $$-$0.175$ & $$-$0.098$ & $0.009$ & $$-$0.017$ & $0.024$ & $$-$0.053$ \\ 
$\Omega^{(d)}_{opd}$ & $0.102$ & $1$ & $0.405$ & $0.125$ & $$-$0.187$ & $$-$0.103$ & $$-$0.618$ & $$-$0.011$ & $0.072$ & $$-$0.202$ & $$-$0.198$ & $0.267$ & $0.070$ & $0.050$ & $0.109$ \\ 
$\Omega^{(o)}_{opd}$ & $0.122$ & $0.405$ & $1$ & $0.129$ & $$-$0.013$ & $$-$0.073$ & $$-$0.565$ & $0.126$ & $0.005$ & $$-$0.234$ & $$-$0.175$ & $0.200$ & $0.051$ & $$-$0.075$ & $0.098$ \\ 
log $x^t_{opd}$ & $0.071$ & $0.125$ & $0.129$ & $1$ & $0.361$ & $0.360$ & $$-$0.194$ & $0.056$ & $0.128$ & $0.048$ & $0.094$ & $0.191$ & $0.104$ & $0.099$ & $0.107$ \\ 
log $x^t_{op}$ & $0.114$ & $$-$0.187$ & $$-$0.013$ & $0.361$ & $1$ & $0.379$ & $0.201$ & $0.157$ & $$-$0.107$ & $0.149$ & $$-$0.042$ & $$-$0.142$ & $$-$0.063$ & $$-$0.127$ & $$-$0.044$ \\ 
log $x^t_{op}$ & $$-$0.018$ & $$-$0.103$ & $$-$0.073$ & $0.360$ & $0.379$ & $1$ & $0.176$ & $0.095$ & $0.590$ & $$-$0.033$ & $0.391$ & $$-$0.151$ & $0.003$ & $$-$0.158$ & $0.103$ \\ 
log $Distance$ & $$-$0.135$ & $$-$0.618$ & $$-$0.565$ & $$-$0.194$ & $0.201$ & $0.176$ & $1$ & $$-$0.022$ & $$-$0.026$ & $0.277$ & $0.248$ & $$-$0.476$ & $$-$0.118$ & $0.0001$ & $$-$0.216$ \\ 
log $gdp^t_o$ & $$-$0.039$ & $$-$0.011$ & $0.126$ & $0.056$ & $0.157$ & $0.095$ & $$-$0.022$ & $1$ & $0.061$ & $$-$0.574$ & $$-$0.001$ & $$-$0.074$ & $$-$0.032$ & $$-$0.095$ & $0.091$ \\ 
log $gdp^t_d$ & $$-$0.070$ & $0.072$ & $0.005$ & $0.128$ & $$-$0.107$ & $0.590$ & $$-$0.026$ & $0.061$ & $1$ & $$-$0.109$ & $$-$0.115$ & $$-$0.075$ & $0.030$ & $$-$0.103$ & $0.124$ \\ 
log $Population_o$ & $$-$0.175$ & $$-$0.202$ & $$-$0.234$ & $0.048$ & $0.149$ & $$-$0.033$ & $0.277$ & $$-$0.574$ & $$-$0.109$ & $1$ & $0.009$ & $$-$0.011$ & $0.036$ & $0.044$ & $$-$0.006$ \\ 
log $Population_d$ & $$-$0.098$ & $$-$0.198$ & $$-$0.175$ & $0.094$ & $$-$0.042$ & $0.391$ & $0.248$ & $$-$0.001$ & $$-$0.115$ & $0.009$ & $1$ & $$-$0.049$ & $0.026$ & $$-$0.025$ & $0.060$ \\ 
$Border_{od}$ & $0.009$ & $0.267$ & $0.200$ & $0.191$ & $$-$0.142$ & $$-$0.151$ & $$-$0.476$ & $$-$0.074$ & $$-$0.075$ & $$-$0.011$ & $$-$0.049$ & $1$ & $0.231$ & $0.123$ & $0.132$ \\ 
$Colony_{od}$  & $$-$0.017$ & $0.070$ & $0.051$ & $0.104$ & $$-$0.063$ & $0.003$ & $$-$0.118$ & $$-$0.032$ & $0.030$ & $0.036$ & $0.026$ & $0.231$ & $1$ & $0.133$ & $0.160$ \\ 
$Language_{od}$ & $0.024$ & $0.050$ & $$-$0.075$ & $0.099$ & $$-$0.127$ & $$-$0.158$ & $0.0001$ & $$-$0.095$ & $$-$0.103$ & $0.044$ & $$-$0.025$ & $0.123$ & $0.133$ & $1$ & $$-$0.100$ \\ 
log$Lang. Proximity_{od}$ & $$-$0.053$ & $0.109$ & $0.098$ & $0.107$ & $$-$0.044$ & $0.103$ & $$-$0.216$ & $0.091$ & $0.124$ & $$-$0.006$ & $0.060$ & $0.132$ & $0.160$ & $$-$0.100$ & $1$ \\ 
\hline \\[-1.8ex] 
\end{tabular} 
\end{adjustbox}
\end{table} 

\begin{table}[!htbp] \centering 
  \caption{Correlation Matrix: Experienced exporters} 
  \label{} 
      \begin{adjustbox}{width=1.2\textwidth,center}
\begin{tabular}{@{\extracolsep{5pt}} cccccccccccccccc} 
\\[-1.8ex]\hline 
\hline \\[-1.8ex] 
 & $\omega^t_{opd}$  & $\Omega^{(d)}_{opd}$ & $\Omega^{(o)}_{opd}$ & log $x^t_{opd}$ & log $x^t_{op}$ & log $x^t_{pd}$ & log $Distance$ & log $gdp^t_o$ & log $gdp^t_d$ & log $Population_o$ & log $Population_d$ & $Border_{od}$ & $Colony_{od}$ & $Language_{od}$ & log$Lang. Proximity_{od}$ \\ 
\hline \\[-1.8ex] 
$\omega^t_{opd}$  & $1$ & $0.047$ & $0.066$ & $0.121$ & $0.205$ & $0.036$ & $$-$0.034$ & $$-$0.052$ & $$-$0.048$ & $$-$0.095$ & $$-$0.072$ & $$-$0.006$ & $$-$0.028$ & $$-$0.044$ & $$-$0.016$ \\ 
$\Omega^{(d)}_{opd}$ & $0.047$ & $1$ & $0.327$ & $0.113$ & $$-$0.178$ & $0.033$ & $$-$0.607$ & $0.004$ & $0.163$ & $$-$0.191$ & $$-$0.135$ & $0.236$ & $0.022$ & $$-$0.026$ & $0.138$ \\ 
$\Omega^{(o)}_{opd}$ & $0.066$ & $0.327$ & $1$ & $0.049$ & $$-$0.035$ & $0.009$ & $$-$0.469$ & $0.128$ & $0.053$ & $$-$0.208$ & $$-$0.106$ & $0.107$ & $$-$0.004$ & $$-$0.071$ & $0.101$ \\ 
log $x^t_{opd}$ & $0.121$ & $0.113$ & $0.049$ & $1$ & $0.417$ & $0.468$ & $$-$0.165$ & $0.103$ & $0.209$ & $0.093$ & $0.136$ & $0.154$ & $0.062$ & $0.050$ & $0.137$ \\ 
log $x^t_{op}$ & $0.205$ & $$-$0.178$ & $$-$0.035$ & $0.417$ & $1$ & $0.303$ & $0.185$ & $0.205$ & $$-$0.138$ & $0.209$ & $$-$0.082$ & $$-$0.131$ & $$-$0.022$ & $$-$0.081$ & $$-$0.013$ \\ 
log $x^t_{pd}$ & $0.036$ & $0.033$ & $0.009$ & $0.468$ & $0.303$ & $1$ & $0.030$ & $0.036$ & $0.469$ & $$-$0.115$ & $0.311$ & $$-$0.017$ & $$-$0.045$ & $$-$0.062$ & $0.109$ \\ 
log $Distance$ & $$-$0.034$ & $$-$0.607$ & $$-$0.469$ & $$-$0.165$ & $0.185$ & $0.030$ & $1$ & $$-$0.082$ & $$-$0.132$ & $0.255$ & $0.186$ & $$-$0.447$ & $$-$0.039$ & $0.028$ & $$-$0.250$ \\ 
log $gdp^t_o$ & $$-$0.052$ & $0.004$ & $0.128$ & $0.103$ & $0.205$ & $0.036$ & $$-$0.082$ & $1$ & $0.025$ & $$-$0.468$ & $$-$0.024$ & $$-$0.082$ & $0.051$ & $$-$0.010$ & $0.097$ \\ 
log $gdp^t_d$ & $$-$0.048$ & $0.163$ & $0.053$ & $0.209$ & $$-$0.138$ & $0.469$ & $$-$0.132$ & $0.025$ & $1$ & $$-$0.085$ & $$-$0.185$ & $0.037$ & $$-$0.018$ & $$-$0.036$ & $0.153$ \\ 
log $Population_o$ & $$-$0.095$ & $$-$0.191$ & $$-$0.208$ & $0.093$ & $0.209$ & $$-$0.115$ & $0.255$ & $$-$0.468$ & $$-$0.085$ & $1$ & $$-$0.017$ & $$-$0.015$ & $0.026$ & $$-$0.007$ & $$-$0.015$ \\ 
log $Population_d$ & $$-$0.072$ & $$-$0.135$ & $$-$0.106$ & $0.136$ & $$-$0.082$ & $0.311$ & $0.186$ & $$-$0.024$ & $$-$0.185$ & $$-$0.017$ & $1$ & $0.004$ & $$-$0.002$ & $0.011$ & $0.062$ \\ 
$Border_{od}$ & $$-$0.006$ & $0.236$ & $0.107$ & $0.154$ & $$-$0.131$ & $$-$0.017$ & $$-$0.447$ & $$-$0.082$ & $0.037$ & $$-$0.015$ & $0.004$ & $1$ & $0.125$ & $0.130$ & $0.137$ \\ 
$Colony_{od}$ & $$-$0.028$ & $0.022$ & $$-$0.004$ & $0.062$ & $$-$0.022$ & $$-$0.045$ & $$-$0.039$ & $0.051$ & $$-$0.018$ & $0.026$ & $$-$0.002$ & $0.125$ & $1$ & $0.332$ & $0.044$ \\ 
$Language_{od}$ & $$-$0.044$ & $$-$0.026$ & $$-$0.071$ & $0.050$ & $$-$0.081$ & $$-$0.062$ & $0.028$ & $$-$0.010$ & $$-$0.036$ & $$-$0.007$ & $0.011$ & $0.130$ & $0.332$ & $1$ & $$-$0.095$ \\ 
log$Lang. Proximity_{od}$ & $$-$0.016$ & $0.138$ & $0.101$ & $0.137$ & $$-$0.013$ & $0.109$ & $$-$0.250$ & $0.097$ & $0.153$ & $$-$0.015$ & $0.062$ & $0.137$ & $0.044$ & $$-$0.095$ & $1$ \\ 
\hline \\[-1.8ex] 
\end{tabular} 
\end{adjustbox}
\end{table} 

\end{landscape}
\restoregeometry
\clearpage

\section*{Appendix E. Bilateral trade volume after two years for primary products, resource-based, low-tech, medium-tech, and high-tech manufactures}
\label{appendixE}

Following \cite{Lall2000}, we exclude "special transactions" such as electric current, cinema film, printed matter, fold, coins, and pets.  
 
\begin{table}[!htbp] \centering 
  \caption{Lall's classification} 
  \label{} 
    \begin{adjustbox}{width=1\textwidth,center}
\begin{tabular}{@{\extracolsep{5pt}}lD{.}{.}{-3} D{.}{.}{-3} D{.}{.}{-3} D{.}{.}{-3} D{.}{.}{-3} } 
\\[-1.8ex]\hline 
\hline \\[-1.8ex] 
 & \multicolumn{5}{c}{\textit{Dependent variable}:~log $x^{t+2}_{opd}$} \\ 
\cline{2-6} 
\\[-1.8ex] & \multicolumn{1}{c}{Primary Product} & \multicolumn{1}{c}{Resource-based Manufactures} & \multicolumn{1}{c}{Low-tech Manufactures} & \multicolumn{1}{c}{Medium-tech Manufactures} & \multicolumn{1}{c}{High-tech Manufactures}\\ 
\hline \\[-1.8ex] 
  $\omega^t_{opd}$  & 0.183^{***} & 0.164^{***} & 0.204^{***} & 0.203^{***} & 0.229^{***} \\ 
  & (0.003) & (0.002) & (0.001) & (0.003) & (0.003) \\ 
  $\Omega^{(d)}_{opd}$ & 0.152^{***} & 0.144^{***} & 0.131^{***} & 0.154^{***} & 0.128^{***} \\ 
  & (0.003) & (0.002) & (0.002) & (0.003) & (0.003) \\ 
  $\Omega^{(o)}_{opd}$ & 0.098^{***} & 0.076^{***} & 0.072^{***} & 0.084^{***} & 0.078^{***} \\ 
  & (0.003) & (0.002) & (0.002) & (0.003) & (0.003) \\ 
  log $x^t_{opd}$ & 0.343^{***} & 0.375^{***} & 0.409^{***} & 0.362^{***} & 0.387^{***} \\ 
  & (0.001) & (0.001) & (0.0005) & (0.001) & (0.001) \\ 
  log $x^t_{op}$ & 0.353^{***} & 0.341^{***} & 0.325^{***} & 0.369^{***} & 0.333^{***} \\ 
  & (0.001) & (0.001) & (0.001) & (0.002) & (0.001) \\ 
  log $x^t_{pd}$ & 0.565^{***} & 0.490^{***} & 0.428^{***} & 0.509^{***} & 0.615^{***} \\ 
  & (0.004) & (0.003) & (0.002) & (0.004) & (0.004) \\ 
  log $Distance$  & -0.421^{***} & -0.439^{***} & -0.495^{***} & -0.544^{***} & -0.477^{***} \\ 
  & (0.004) & (0.003) & (0.002) & (0.004) & (0.004) \\ 
  log $gdp^t_o$ & 0.103^{***} & 0.082^{***} & 0.166^{***} & 0.132^{***} & 0.296^{***} \\ 
  & (0.003) & (0.002) & (0.002) & (0.004) & (0.004) \\ 
  log $gdp^t_d$ & 0.151^{***} & 0.184^{***} & 0.292^{***} & 0.164^{***} & 0.271^{***} \\ 
  & (0.004) & (0.002) & (0.002) & (0.004) & (0.004) \\ 
  log $Population_o$ & 0.326^{***} & 0.419^{***} & 0.482^{***} & 0.511^{***} & 0.517^{***} \\ 
  & (0.003) & (0.002) & (0.002) & (0.004) & (0.004) \\ 
  log $Population_d$ & 0.298^{***} & 0.323^{***} & 0.339^{***} & 0.349^{***} & 0.379^{***} \\ 
  & (0.003) & (0.002) & (0.002) & (0.003) & (0.003) \\ 
  $Border_{od}$ & 0.741^{***} & 0.754^{***} & 0.668^{***} & 0.733^{***} & 0.664^{***} \\ 
  & (0.010) & (0.007) & (0.006) & (0.011) & (0.011) \\ 
  $Colony_{od}$ & -0.112^{***} & -0.016^{**} & 0.077^{***} & 0.037^{***} & 0.182^{***} \\ 
  & (0.011) & (0.007) & (0.006) & (0.011) & (0.012) \\ 
  $Language_{od}$ & 0.395^{***} & 0.479^{***} & 0.553^{***} & 0.556^{***} & 0.619^{***} \\ 
  & (0.008) & (0.005) & (0.004) & (0.008) & (0.008) \\ 
  log$Lang. Proximity_{od}$ & 0.043^{***} & 0.038^{***} & 0.031^{***} & 0.036^{***} & 0.021^{***} \\ 
  & (0.003) & (0.002) & (0.001) & (0.003) & (0.003) \\ 
  Constant & 0.653^{***} & 0.484^{***} & 0.332^{***} & 0.274^{***} & 0.529^{***} \\ 
  & (0.020) & (0.015) & (0.012) & (0.025) & (0.024) \\ 
 \hline \\[-1.8ex] 
Observations & \multicolumn{1}{c}{1,127,670} & \multicolumn{1}{c}{2,241,432} & \multicolumn{1}{c}{3,314,246} & \multicolumn{1}{c}{1,110,342} & \multicolumn{1}{c}{1,049,765} \\ 
Adjusted R$^{2}$ & \multicolumn{1}{c}{0.399} & \multicolumn{1}{c}{0.446} & \multicolumn{1}{c}{0.527} & \multicolumn{1}{c}{0.471} & \multicolumn{1}{c}{0.565} \\ 
Residual Std. Error & \multicolumn{1}{c}{2.874} & \multicolumn{1}{c}{2.658} & \multicolumn{1}{c}{2.371} & \multicolumn{1}{c}{2.689} & \multicolumn{1}{c}{2.477} \\ 
\hline 
\hline \\[-1.8ex] 
\textit{Note:}& \multicolumn{5}{r}{$^{*}$p$<$0.1; $^{**}$p$<$0.05; $^{***}$p$<$0.01} \\ 
\end{tabular} 
\end{adjustbox}
\end{table} 
\clearpage

\end{document}